\documentclass[11pt]{article}
\usepackage{eurosym}
\usepackage{natbib}
\usepackage{amssymb}
\usepackage{amsmath}
\usepackage[scr=rsfs]{mathalpha}
\usepackage{bm}
\usepackage{graphicx}
\usepackage{epstopdf}
\usepackage[singlespacing]{setspace}
\usepackage[bottom,multiple]{footmisc}
\usepackage[justification=centering]{caption}
\usepackage{subcaption}
\usepackage{setspace}
\usepackage{color}
\usepackage{ragged2e}
\usepackage{bigstrut}
\usepackage{lscape}
\usepackage{arydshln}
\usepackage{booktabs}                                         
\usepackage{multirow}                                          
\usepackage{comment}

\RequirePackage[breaklinks, colorlinks,citecolor=black,urlcolor=black,linkcolor=black]{hyperref}

\newcommand{\qed}{\nobreak \ifvmode \relax \else
      \ifdim\lastskip<1.5em \hskip-\lastskip
      \hskip1.5em plus0em minus0.5em \fi \nobreak
      \vrule height0.3em width0.5em depth0.25em\fi}
\newcommand{\vm}{\vspace{.1in}}

\def\toprule{\hline\hline}
\def\midrule{\hline}
\def\bottomrule{\hline\hline}
\def\tablenotes{\\ \noindent \justifying \small}

\def\toprule{\hline\hline\addlinespace[0.2cm]}
\def\midrule{\addlinespace[0.1cm]\hline\addlinespace[0.1cm]}
\def\bottomrule{\addlinespace[0.2cm]\hline\hline}
\def\tablenotes{\\ \vm \noindent \justifying \footnotesize Notes: }

\renewcommand{\today}{\ifcase \month \or January \or February \or March \or %
April \or May \or June \or July \or August \or September \or October \or November \or %
December \fi \number \year}

\usepackage{titling}

\begin{document}

\title{Branding through responsibility: the advertising impact of CSR activities in the Korean instant noodles market\thanks{This paper previously circulated under the title `Does it pay to be a reputable company? Insights from the instant noodles market.' We are grateful to the seminar participants at Sogang University, the 2024 Annual Meeting of Korea's Allied Economic Association (KAEA), and the 2024 Annual Conference of the International Association of Applied Econometrics (IAAE) for their valuable feedback and suggestions. This research was supported by the BK21 FOUR program, funded by the Ministry of Education and the National Research Foundation of South Korea, as well as the Dr. Delia Koo Faculty Endowment Award, granted by the Asian Studies Center of Michigan State University.}}
\author{Youngjin Hong\thanks{%
Department of Economics, University of Michigan, 258 Lorch Hall 611 Tappan Ave., Ann Arbor, MI 48109, E-mail: \texttt{yjinhong@umich.edu}.} \and
In Kyung Kim\thanks{%
Corresponding author, Department of Economics, Sogang University, 35 Baekbeom-ro Mapo-gu, Seoul 04107, South Korea, E-mail: \texttt{inkim@sogang.ac.kr}.} \and %
Kyoo il Kim\thanks{%
Department of Economics, Michigan State University, 486 West Circle Drive, East Lansing, MI 48824, E-mail: \texttt{kyookim@msu.edu}.}}

\maketitle

\begin{abstract}
This paper empirically examines the extent to which a favorable view of a firm, shaped by its social contributions, influences consumer choices and firm sales. Using a favorability rating that reflects media exposure of each firm's corporate social responsibility (CSR) activities in the Korean instant noodles market during the 2010s, we find evidence that improvements in the corporate image of Ottogi --- one of the country’s largest instant noodle producers --- positively affected consumer utility for the firm’s products. Notably, Ottogi’s annual sales of its major brands increased by an average of 23.7 million packages, or 6.7\%, as a result of CSR activities and the associated rise in consumer favorability. This effect is comparable in magnitude to that of a nearly 60\% increase in advertising spending. Our findings suggest that CSR can foster firm growth by boosting product sales.
\bigskip

\end{abstract}

\doublespace

\clearpage

\section{Introduction}

 An increasing number of firms are now explicitly prioritizing corporate social responsibility (CSR) as a central element of their business operations. Firms engaged in CSR initiatives contribute to societal well-being by assisting local communities through charity donations, promoting equality and workplace rights, and actively reducing negative externalities beyond the legal minimum.\footnote{The following website provides 20 such examples: \\ \url{https://odysseyteams.com/corporate-philanthropy-examples/}} Consistent commitments to CSR not only generate social impacts but also positively influence corporate images as perceived by consumers. This enhanced image may play a pivotal role in shaping a firm's performance in the final goods market. This is because in making purchasing decisions, consumers often evaluate not just the inherent attributes of a product but also the associated corporate image, frequently placing higher value on products linked with a positive corporate image. For instance, 55\% of the respondents to a global survey conducted by NielsenIQ in 2014 said that they are willing to pay more for goods and services from socially responsible firms.\footnote{\url{https://nielseniq.com/global/en/insights/report/2014/doing-well-by-doing-good/}} Despite growing attention and investment in CSR activities by firms, however, the extent to which this investment in CSR, and the resultant positive corporate image, translates into additional consumer demand and subsequent revenue increase remains unclear.

 In this paper, we examine whether and to what extent a favorable view of a firm, shaped by its social contributions, affects consumer choices and firm sales. We study this issue in the context of the South Korean instant noodles (\textit{ramen}) market, where Ottogi, one of the major \textit{ramen} makers, established a distinguished corporate image through various CSR activities in the 2010s. This period coincides with a gradual increase in the firm's revenue share in the instant noodles market. Unlike previous works (e.g., \citealp{auger2003will, inoue2017predicting, nickerson2022impact}), we consider strategic interactions among firms in the market and quantify the extent to which a firm’s enhanced favorability, driven by CSR, affects not only its own sales and revenues but also those of its competitors. Furthermore, our counterfactual analysis presents an advertising strategy that could achieve revenues equivalent to those attained through CSR initiatives.

 Based on the theoretical literature on CSR (e.g., \citealp{bagnoli2003selling, baron2009positive}), we consider a consumer utility specification that accounts for firms' CSR efforts. To this end, we carefully construct a favorability rating that reflects media exposure of a firm's CSR activities. We then estimate a nested logit demand model using instant noodle sales data from NielsenIQ, which contains monthly sales and price information for each product from the four major South Korean instant noodle producers between August 2010 and December 2019. Our estimation results provide evidence that Ottogi's CSR performance and the resulting improvement in its corporate image positively affected consumer utility for the firm's products, above and beyond the impact of advertising expenditure.

 Next, we measure the sales effect of corporate image improvement through a series of counterfactual analyses. Specifically, using the demand estimates, we first examine what the firm's sales revenues would have been during the 2010s had the change in Ottogi's image followed a pattern similar to that of its rivals. Comparing these counterfactual sales with the observed figures, we find that the favorable perception of Ottogi increased the firm's sales by 6.7\% (approximately 2 million more packages sold per month) during the sample period. Ottogi's benefit from its image improvement is equivalent to a 6.5\% revenue increase (about 1.3 billion more Korean Won per month, or equivalently 1.2 million more US dollars per month).\footnote{The average exchange rate during the 2010s, 0.89 US dollars per one thousand Korean Won, is used throughout the paper.} We also find that this sales effect is comparable in magnitude to that of a nearly 60\% increase in advertising expenditure. These findings illustrate that the steady increase in Ottogi's market share in the \textit{ramen} market during the 2010s — from below 10\% in 2010 to over 20\% in 2019 — can largely be attributed to the positive impact of the firm's diverse CSR activities on its corporate image.

 Conversely, Ottogi's three competitors experienced revenue declines of 1.3\%--2.1\%, as their sales dropped by more than 1\% due to Ottogi’s enhanced image. Moreover, they had to offer lower prices to compensate for consumers' diminished appetite for their products, which stemmed from relatively weaker corporate images. In contrast, Ottogi leveraged consumers' appreciation for its CSR activities to charge higher prices for its products. Our findings are robust across different model specifications and remain qualitatively unchanged when accounting for the potential endogeneity of advertising.

 Previous studies have shown that consumers tend to avoid purchasing products from firms that fail to fulfill their social responsibilities, such as abusing power over business partners (Kim and Kim, 2022), treating employees unfairly \citep{skarlicki2004third}, or harming the environment \citep{hoffmann2018under}. While consumers punish such firms with tarnished reputations by boycotting their products, our findings suggest they tend to favor products from reputable firms and are willing to pay more for them. Therefore, positive publicity helps a firm not only perform better in the stock market \citep{flammer2015does} and attract more productive employees \citep{hedblom2019toward}, but also achieve higher sales and revenues. Given that a favorable corporate image facilitates firm growth through various channels, firms may need to enhance their reputations by engaging in CSR activities such as adhering to ethical and sustainable management and creating a favorable working environment.

 The remainder of the paper is structured as follows: In Section 2, we review related works and delineate our contribution to the literature. Section 3 details the South Korean \textit{ramen} market and describes the data. After introducing and estimating the instant noodle demand model in Section 4, we conduct counterfactual analyses and quantify the sales effect of corporate image improvement in Section 5. We discuss managerial implications and conclude in Section 6.

\section{Related literature}

 Theoretical literature on CSR has investigated the relationship between firms' CSR efforts and market structure, as well as the resulting economic outcomes (e.g., \citealp{bagnoli2003selling, baron2009positive, calveras2016role, schinkel2024corporate}). In these studies, consumers are assumed to incorporate firms' CSR provisions or efforts into their utilities when making purchase decisions. Theoretical implications therefore depend not only on whether consumers value firms’ CSR efforts positively, but also on the degree of this valuation --- both of which are important empirical questions. We examine whether, and to what extent, a favorable view of a firm induced by CSR activities affects actual consumer choices.

 This paper relates to the empirical literature on the effects of CSR on firm performance. One strand of studies has examined CSR's impact on financial market outcomes. \cite{flammer2015does} utilized a regression discontinuity design to find that adopting CSR proposals can lead to an increase in shareholder value and enhance accounting performance. \cite{peters2009some} and \cite{jeong2018permanency} suggested that the effects of CSR on financial performance and firm value might be cumulative and realized over the long term, rather than transitory and short-lived. Another body of research has investigated CSR's effects on labor market outcomes. \cite{hedblom2019toward} presented evidence that CSR may increase the number of applicants and attract more productive employees. Conversely, \cite{list2021corporate} suggested that CSR usage might backfire, increasing employee misbehavior, which in turn weakens the firm's profitability and performance.\footnote{Additionally, some previous works have documented a positive relationship between CSR and firms' investment decisions in green technologies. While \cite{hao2022corporate} used third-party CSR ratings to capture firm-level CSR performance, \cite{mbanyele2022corporate} relied on the staggered adoption of CSR disclosure.} Our paper contributes to this literature by providing empirical evidence of CSR's effect on product market outcomes.

 Another strand of empirical literature has examined the impact of CSR on consumers' purchase intentions and behavioral loyalty, as well as its impact on product sales. Several survey-based studies (e.g., \citealp{auger2003will, mohr2005effects, martinez2013csr}) suggested that consumers generally form a positive association between CSR and products or firms, leading to higher purchase intention and loyalty towards these firms.\footnote{According to \cite{auger2003will}, consumers tend to rate products associated with ethical and social responsibility highly. \cite{mohr2005effects} argued that CSR in the environmental domain might more strongly affect a consumer's willingness to purchase than price. \cite{martinez2013csr} showed, in the context of the Spanish hotel industry, that perceived CSR indirectly and positively affects customer loyalty.} On the other hand, in the context of a professional sports team, \cite{inoue2017predicting} utilized both attitudinal surveys and actual attendance data to show that the contribution of CSR initiatives to behavioral loyalty may not be significant. \cite{van2021does} showed that investments in CSR activities boost sales of newly introduced sustainable products, and \cite{nickerson2022impact} found that different categories of CSR activities have heterogeneous impacts on brand sales.

 We contribute to the literature by presenting a more comprehensive picture of the effects of CSR on the entire industry, rather than just on specific brands or products. Spanning nearly ten years of monthly sales data, our study quantifies the extent to which a favorable view of a firm, shaped by its CSR activities, influences the prices and sales of both the firm and its competitors. To achieve this, we adopt a discrete choice model of differentiated product demand and capture product substitution arising from Ottogi’s distinctive level of CSR activities and the resulting improvement in its corporate image. Additionally, our counterfactual analyses allow us to propose a comparable advertising strategy that could generate revenues equivalent to those achieved through CSR initiatives. We therefore complement previous studies on CSR that rely on survey-based methods and reduced-form analyses, providing a richer examination of the broader effects of CSR on product market outcomes.

 While our study focuses on the impact of improved corporate image on consumer choices and firm sales, an alternative body of literature has explored the consequences of tarnished corporate images. \cite{bachmann2023firms}, studying the impact of Volkswagen's emissions scandal in the German automobile industry, demonstrated that the firm's negative reputation could have negative spillover effects on the sales of other firms sharing the group identity. \cite{sharpe2021sales} found a negative correlation between corporate social irresponsibility and sales and a mitigating role for advertising. This strand of research has also investigated the effects of boycotts stemming from worsened images of a firm or country.\footnote{\cite{kim2021corporate} presented empirical evidence on the consequences of a boycott against a South Korean dairy firm triggered by its unethical management practices. Studies examining a surrogate boycott, a movement against a foreign country triggered by historical animosity and international politics conflict, include \cite{sun2020consumer}, \cite{clerides2015national}, and \cite{kim2022no}. Also, \cite{kim2023country} showed that the worsened image of China caused sales of Chinese beers in South Korea to decrease during the Covid-19 pandemic.} Contrary to these previous works, our study delves into the effects of an improved corporate image, fostered through diverse CSR initiatives, on consumer choices and sales performance.

\section{Background information and data}

 \subsection*{Instant noodles market in South Korea}

 In South Korea, the first instant noodle product was introduced by Samyang Foods (hereafter, Samyang) in 1963. Since then, instant noodles, known for their quick preparation and affordability, have become a staple dietary choice among South Koreans. With a monthly per capita consumption of over five packages, instant noodles account for 2.8\% of daily calorie intakes in South Korea, making them the third most significant source of energy, just behind white rice (32.7\%) and pork (4.6\%).\footnote{Source: 2010 Korea National Health and Nutrition Examination Survey (\url{https://www.mohw.go.kr/board.es?mid=a10411010100&bid=0019}). The survey data from the Ministry of Health and Welfare in 2010 and our observed data from NielsenIQ yield a comparable average number of monthly per capita consumption.} Consequently, the \textit{ramen} category has become one of the largest segments of the processed food industry in South Korea.

 Due to the high popularity and heavy reliance on instant noodle products among Koreans, successive Korean governments have consistently sought to keep \textit{ramen} prices under control since designating them as administratively regulated items in the late 1970s. Although government approval is no longer required before raising prices, instant noodle manufacturers still face pressure to restrain price increases, particularly during periods of inflation such as the global financial crisis in the late 2000s and the COVID-19 pandemic in the early 2020s. We account for this \textit{de facto} price control in our counterfactual analysis in Section~5.\footnote{\cite{kim2025collusion} provided a detailed description of price regulation in this industry and analyzed its implications for social welfare.}

\begin{figure}[!t]
    \centering
    \caption{CSR activities, corporate image, and sales shares during the 2010s}
    \begin{subfigure}[b]{0.65\textwidth}
    \caption{Sales shares}
    \includegraphics[width=\textwidth]{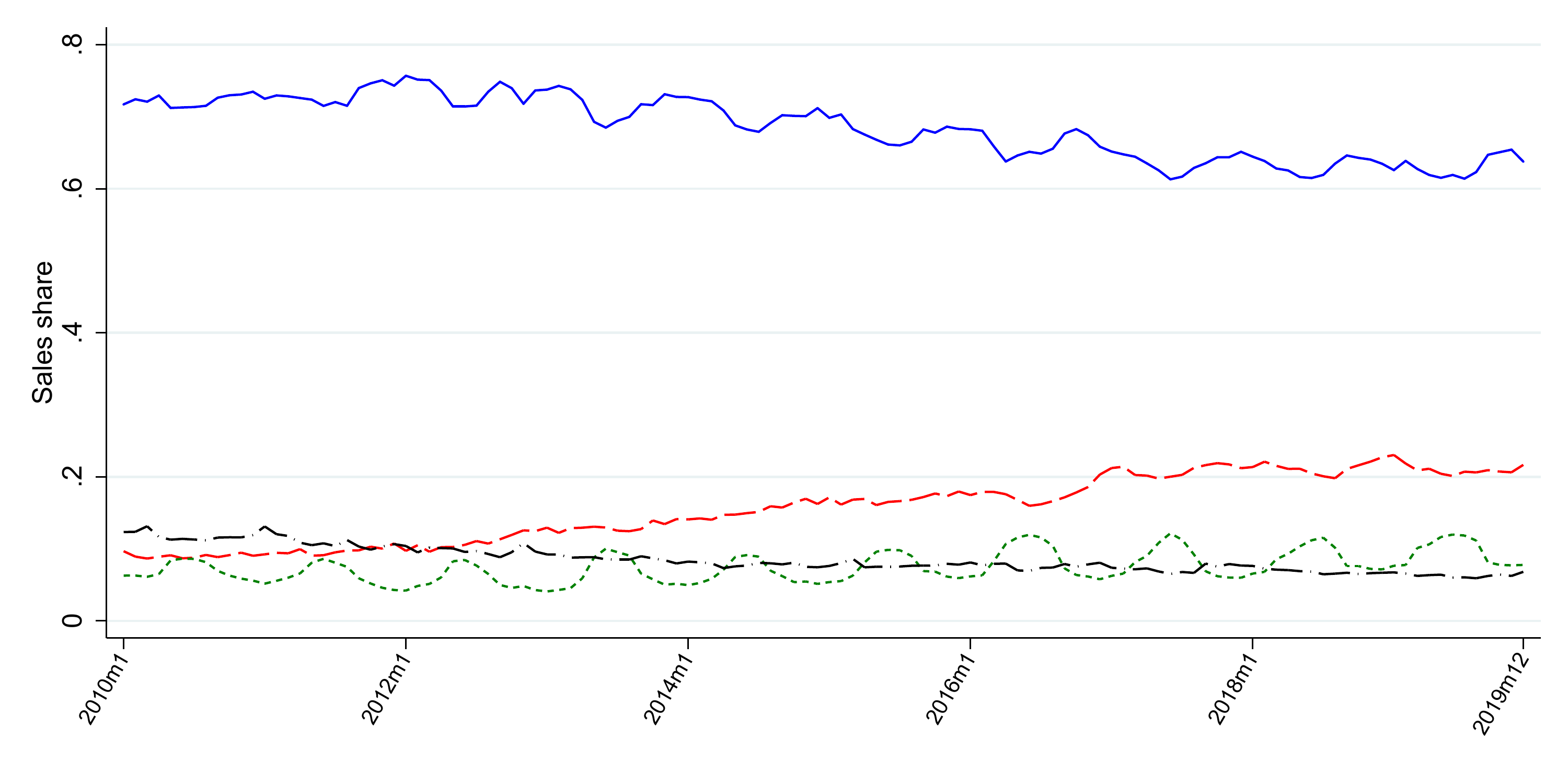}
    \end{subfigure}
    \begin{subfigure}[b]{0.65\textwidth}
    \vspace{.1in}
    \caption{News articles on CSR activities}
    \includegraphics[width=\textwidth]{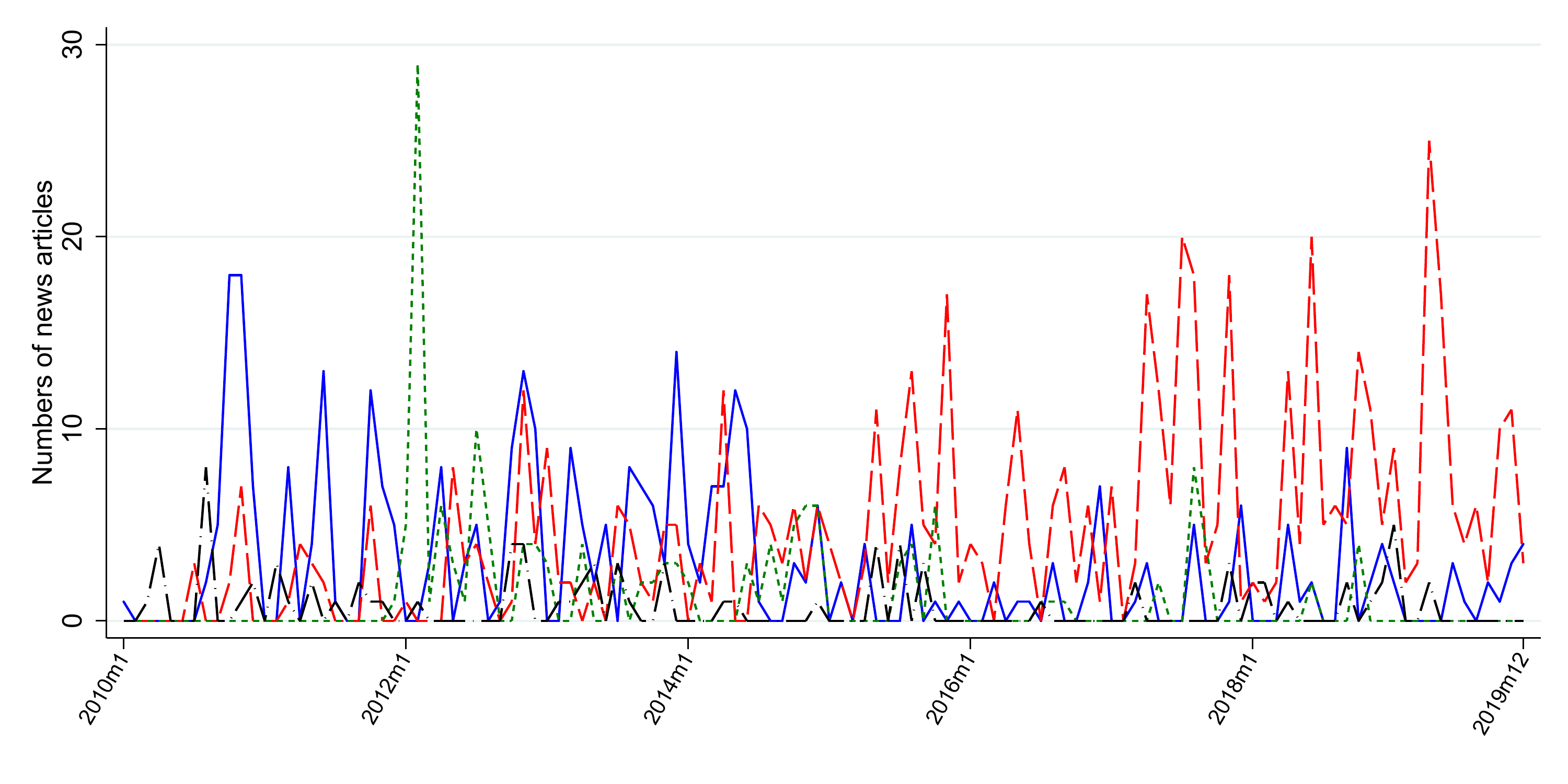}
    \end{subfigure}
    \begin{subfigure}[b]{0.65\textwidth}
    \vspace{.1in}
    \caption{Corporate image scores}
    \includegraphics[width=\textwidth]{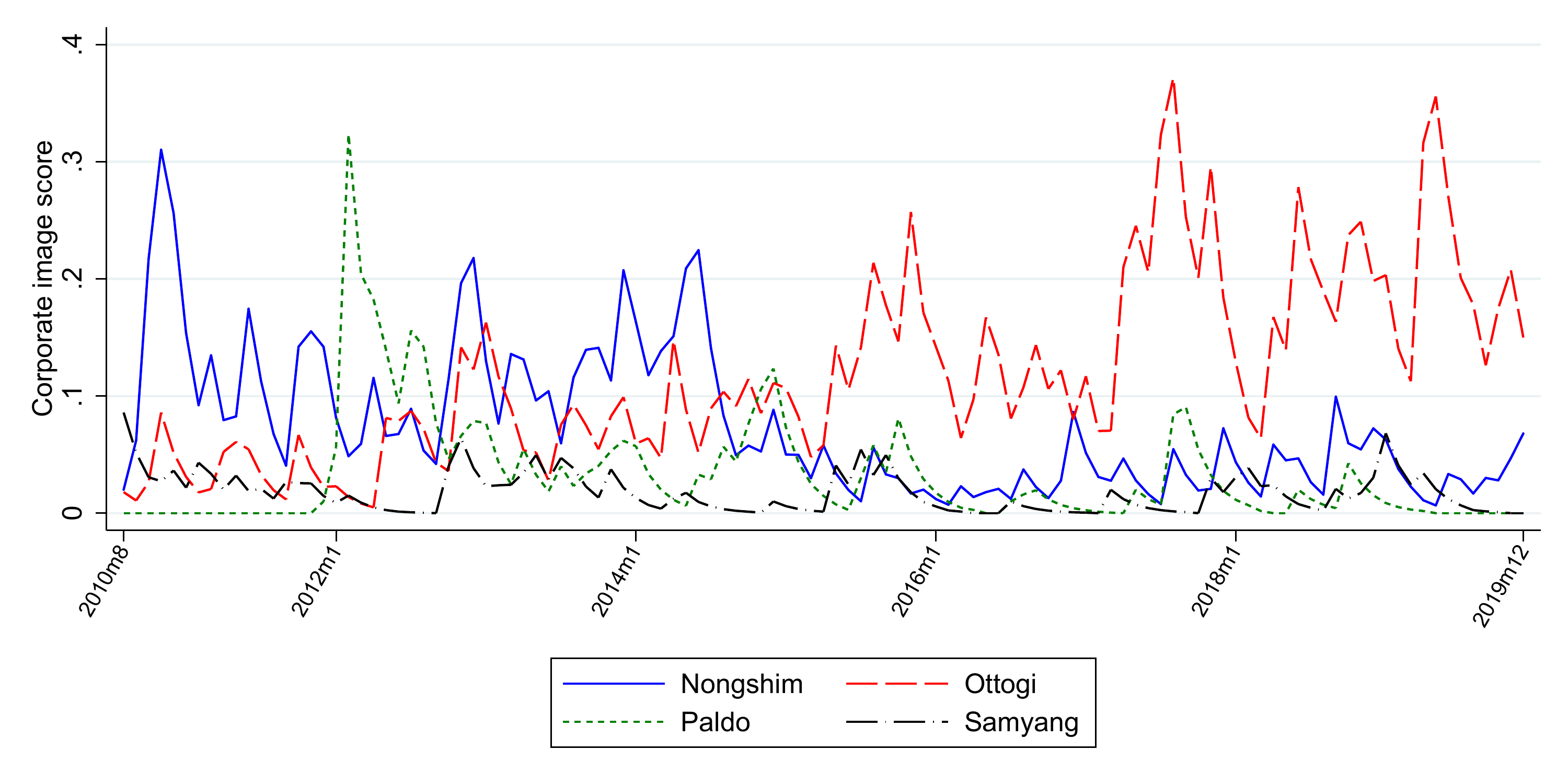}
    \end{subfigure}
\label{fig: share and image}%
\tablenotes The top and middle panels present the monthly sales shares and the number of news articles on CSR activities for the four firms during the 2010s, respectively. Only products sold throughout the entire period are included in the calculation of sales shares. The bottom panel displays each firm’s monthly corporate image score over the sample period. Source: NielsenIQ and Bigkinds.
\end{figure}

 Four major firms --- Nongshim, Ottogi, Samyang, and Paldo --- have dominated the industry, accounting for over 95\% of the market sales over the past two decades. Nongshim, the market leader, generated a total sales of 1.37 trillion Korean Won in 2010, constituting three-quarters of the four firms' total sales (1.85 trillion Korean Won). Meanwhile, Ottogi, with a sales revenue of 156 billion Korean Won, or approximately 8\% of the total sales of the four firms, gradually increased its sales to 371 billion Korean Won by 2019, taking approximately 20\% of the market revenue. The top panel of Figure \ref{fig: share and image} depicts the dynamics of the sales shares of the four firms between 2010 and 2019, showing that Ottogi solidified its position as the second largest by the late 2010s.

 During the 2010s, Ottogi actively engaged in various CSR activities, such as contributing to local charities and hospitals and establishing educational foundations to support tuition and scholarly projects. Additionally, unlike many other firms that resorted to legal strategies to reduce their inheritance taxes, Ottogi diligently paid its dues amounting to approximately 150 billion Korean Won (equivalent to 134 million US dollars). The company also maintained the lowest share of temporary workers in the processed food sector, thanks to its distinctive full-time employment policy. Ottogi's corporate culture emphasizing social contribution, instilled by the late chairman and founder Mr.~Taeho Ham, was notably ahead of its time; most Korean firms have only recently begun to prioritize ESG management.\footnote{In January 2021, the Korean Financial Services Commission announced mandatory ESG disclosures for large corporations starting in 2025.}

 The middle panel of Figure~\ref{fig: share and image} shows that an increasing number of news articles highlighted Ottogi's CSR efforts throughout the 2010s, helping consumers recognize the firm’s strong CSR performance, which consistently exceeded both regulatory requirements and legal standards. In Korea, the firm's highly positive image is often associated with the popular internet meme ``\textit{God}ttogi,'' a portmanteau of ``God'' (used colloquially to mean ``divine'') and ``Ottogi.''\footnote{See the following article for an example of consumers’ recognition of Ottogi’s various social contribution activities during the 2010s: \url{https://www.mk.co.kr/news/business/7882681}.} As Ottogi established a distinguished corporate image among consumers, Mr.~Ham Youngjoon, Chief Executive Officer of Ottogi, was invited to a dinner reception with national business leaders at the Blue House, the executive office of the President of South Korea, in July 2017. According to the Blue House, Ottogi was the only mid-sized and the only food company represented at the event, recognized for its exemplary business practices, including fostering cooperative relationships, championing workers’ rights, and cultivating a positive corporate image among consumers.

 \subsection*{Construction of a favorability rating}\label{section: proxy}

 In this study, we construct a favorability rating based on media exposure of a firm's CSR activities. To this end, we exploit news article data sourced from Bigkinds, a South Korean data analytics platform that specializes in comprehensive media analysis by aggregating content from a wide range of news sources.\footnote{\url{https://www.bigkinds.or.kr/}} Bigkinds offers both a historical archive of news articles and advanced search engine tools that enable precise collection of articles containing specific keywords. Using these tools, we first collected news articles about each firm containing several keywords indicative of CSR activities leading to a positive corporate reputation: donation, contribution, benefaction, community service, volunteer, support, endorsement, ESG, foundation, social (enterprise), etc. After manually reviewing each selected article to filter out irrelevant content, we counted only those articles that highlight CSR activities for each firm and time period (year-month). Further details regarding the collection of news articles and the exclusion procedure, crucial for avoiding potential reverse causality, are provided in Appendix A.

 Similar to the case of advertising (e.g., \citealp{clarke1976econometric, bagwell2007economic}), favorable views toward a firm resulting from its CSR activities may persist in consumers' memory. To account for such lasting effects on consumer preferences, we construct the corporate image score ($imgscore$) of firm $f$ at time $t$ and cumulative advertising expenditures ($cumadv$) of product $j$ at time $t$ as follows:
    \begin{equation}\label{eqn: image score}
    imgscore_{ft} = \sum_{n=0}^{k}(1-\delta)^{n}news_{f,t-n}
    \end{equation}
    \begin{equation}\label{eqn: ad expenditure}
    cumadv_{jt} = \sum_{n=0}^{k}(1-\delta)^{n}adv_{j,t-n},
    \end{equation}
 where $news$ is the firm-level monthly CSR-related news article count and $adv$ is the brand-level monthly advertising expenditure. $k$ represents the maximum time horizon to accumulate the variables, and $\delta$ is the geometric depreciation rate.

 The bottom panel of Figure~\ref{fig: share and image}, which presents the trend in each firm's corporate image score~\eqref{eqn: image score} with $k=6$ and $\delta=0.4$, shows that, despite some fluctuations, Ottogi’s favorability rating exhibited a consistent upward trend throughout the 2010s. In contrast, the other three firms showed either a stable or declining trend since the mid-2010s. The improvement in Ottogi's image relative to its rivals coincides with a gradual increase in the firm's sales share in the instant noodles market (as shown in the top panel of the figure), suggesting a positive impact of the former on the latter. While taking $k=6$ and $\delta=0.4$ as our baseline specification, we check the robustness of our empirical results by considering different values for $k$ and $\delta$. We also adopt a linear decaying pattern, as described in Appendix B, and confirm that the results are consistent with those from our baseline specification. Unless otherwise stated, the terms `ad spending' or `ad expenditures' hereafter refer to the baseline cumulative measure.

 \subsection*{Data}\label{section: data}

 In this study, we define a product as a combination of a brand and a package type, namely, either a packet or a cup. For example, Shin Ramyun, Nongshim's flagship brand and the most popular instant noodle brand in South Korea, is available in both formats. Therefore, we consider each packaging format of the brand as a separate product. A market is defined as a region at a given time, with the potential market size assumed to be 10 times the market population. This implies that the maximum per capita consumption of instant noodles could be 10 packages per month.\footnote{Our empirical findings, available from the authors upon request, are robust to variations in the potential market size (7, 15, or 20 packages per month). Additionally, the monthly population data for each of the six regions are obtained from the Korea National Statistical Office: \url{https://kosis.kr/statHtml/statHtml.do?orgId=101&tblId=DT_1B040A3}.} With this market definition, the average market share of the outside option in our sample is approximately 65\%, a figure we consider reasonable.

 The empirical analysis in this study utilizes sales and price information of instant noodles in South Korea from August 2010 to December 2019. The data, sourced from NielsenIQ, provide monthly package prices and sales volumes (number of packages sold) of all instant noodles products offered by the four major firms in each of the six regions of South Korea.\footnote{Figure \ref{fig: market nielsen} in Appendix~D shows how NielsenIQ sorts the country into six regions.} We adjust prices for inflation by dividing them by the monthly Consumer Price Index (CPI) excluding food and energy ($2020=100$). Additionally, we obtain monthly nationwide advertising expenditures for each brand from 2010 to 2019 from Nielsen Media.

 In this study, we focus on the 30 best-selling products in our empirical analysis. 17 products are owned by Nongshim, seven by Ottogi, four by Samyang, and two by Paldo. Their combined revenue accounts for approximately 80\% of the market revenue during the sample period. The products are categorized into four groups based on the type of soup and temperature: red soup (19 products), white soup (three products), soupless (six products), and cold noodles (two products). Red soup, characterized by its red-colored broth, represents the most traditional type of instant noodles in South Korea. White soup, which gained popularity in the early 2010s, is known for its clear, white-colored broth. Soupless noodles, served without broth and mixed with sauce, and cold noodles, meant to be consumed cold, exhibit clear seasonality in sales, being more popular in summer than in winter.

 Our final sample consists of 18,960 observations from 678 markets ($6$ regions $\times$ $113$ year-months). Table \ref{tab: summary stat} presents descriptive statistics of the sample by product group in the upper panel and by firm in the lower panel. According to Panel A, the red soup group has the highest average sales and revenue: 1,257 thousand packages and 936 million Korean Won, respectively. Consequently, the market share --- the ratio of a product's sales volume to the market size --- is the largest on average for red soup products (1.52\%). Despite being the most popular, red soup products are the cheapest, selling for much lower prices (840 Korean Won on average, or approximately three-quarters of a dollar) compared to other groups, which range from 950 to 1,020 Korean Won, or equivalently 0.85 to 0.91 US dollars on average. Advertising expenditures also show large dispersion within and across product categories.

\begin{table}[!t]
\scriptsize
  \centering
  \caption{Summary statistics}
    \begin{tabular}{lrrrrrrrrrrr}
    \hline
    \hline
          &       &       &       &       &       &       &       &       &       &       &  \bigstrut[t]\\
          & \multicolumn{11}{c}{\textit{Panel A: By product group}}              \bigstrut[b]\\
\cline{2-12}          & \multicolumn{2}{c}{Red soup (19)} & \multicolumn{1}{c}{} & \multicolumn{2}{c}{White soup (3)} & \multicolumn{1}{c}{} & \multicolumn{2}{c}{Soupless (6)} & \multicolumn{1}{c}{} & \multicolumn{2}{c}{Cold noodles (2)}  \bigstrut\\
\cline{2-3}\cline{5-6}\cline{8-9}\cline{11-12}    Variable & \multicolumn{1}{c}{Avg.} & \multicolumn{1}{c}{Std. Dev.} & \multicolumn{1}{c}{} & \multicolumn{1}{c}{Avg.} & \multicolumn{1}{c}{Std. Dev.} & \multicolumn{1}{c}{} & \multicolumn{1}{c}{Avg.} & \multicolumn{1}{c}{Std. Dev.} & \multicolumn{1}{c}{} & \multicolumn{1}{c}{Avg.} & \multicolumn{1}{c}{Std. Dev.} \bigstrut\\
    \hline
    Sales volume  & 1256.99 & 1686.82 &       & 318.91 & 338.62 &       & 799.22 & 917.69 &       & 532.99 & 743.64 \bigstrut[t]\\
    Sales revenue & 936.38 & 1164.11 &       & 302.21 & 312.59 &       & 746.46 & 765.13 &       & 434.07 & 550.74 \\
    Market share & 1.52  & 1.77  &       & 0.38  & 0.31  &       & 0.95  & 0.89  &       & 0.61  & 0.74 \\
    Price & 0.84  & 0.19  &       & 0.95  & 0.06  &       & 1.02  & 0.19  &       & 1.00  & 0.25 \\
    Ad expenditure & 394.02 & 807.07 &       & 43.38 & 222.26 &       & 185.44 & 423.70 &       & 464.95 & 594.22 \\
    Image score & 8.40  & 7.27  &       & 5.67  & 5.85  &       & 6.18  & 6.60  &       & 5.36  & 5.90 \\
    \hline
          &       &       &       &       &       &       &       &       &       &       &  \bigstrut[t]\\
          & \multicolumn{11}{c}{\textit{Panel B: By firm}}   \bigstrut[b]\\
\cline{2-12}          & \multicolumn{2}{c}{Ottogi (7)} & \multicolumn{1}{c}{} & \multicolumn{2}{c}{Nongshim (17)} & \multicolumn{1}{c}{} & \multicolumn{2}{c}{Samyang (4)} & \multicolumn{1}{c}{} & \multicolumn{2}{c}{Paldo (2)}  \bigstrut\\
\cline{2-3}\cline{5-6}\cline{8-9}\cline{11-12}    Variable & \multicolumn{1}{c}{Avg.} & \multicolumn{1}{c}{Std. Dev.} & \multicolumn{1}{c}{} & \multicolumn{1}{c}{Avg.} & \multicolumn{1}{c}{Std. Dev.} & \multicolumn{1}{c}{} & \multicolumn{1}{c}{Avg.} & \multicolumn{1}{c}{Std. Dev.} & \multicolumn{1}{c}{} & \multicolumn{1}{c}{Avg.} & \multicolumn{1}{c}{Std. Dev.} \bigstrut\\
    \hline
    Sales volume & 795.18 & 1094.08 & \multicolumn{1}{c}{} & 1173.71 & 1711.65 & \multicolumn{1}{c}{} & 883.86 & 956.34 & \multicolumn{1}{c}{} & 790.69 & 662.80 \bigstrut[t]\\
    Sales revenue & 544.80 & 612.93 & \multicolumn{1}{c}{} & 932.57 & 1226.73 & \multicolumn{1}{c}{} & 701.08 & 643.00 & \multicolumn{1}{c}{} & 689.06 & 501.08 \\
    Market share & 0.94  & 1.05  &       & 1.42  & 1.81  &       & 1.08  & 1.08  &       & 0.95  & 0.60 \\
    Price & 0.83  & 0.25  &       & 0.92  & 0.16  &       & 0.93  & 0.26  &       & 0.90  & 0.13 \\
    Ad expenditure & 505.82 & 953.60 &       & 319.99 & 669.33 &       & 30.01 & 181.26 &       & 326.07 & 471.71 \\
    Image score & 12.63 & 8.16  &       & 7.31  & 6.07  &       & 1.64  & 1.64  &       & 3.29  & 4.88 \\
    \bottomrule
    \end{tabular}%
  \label{tab: summary stat}%
\tablenotes The table presents descriptive statistics of the sample by product group (top panel) and by firm (bottom panel). Sales volume is reported in thousands of packages, while sales revenue and advertising expenditure are in millions of Korean Won. Prices are in thousands of Korean Won. The numbers in parentheses indicate the number of products. The corporate image score and advertising expenditure are calculated based on equations \eqref{eqn: image score} and \eqref{eqn: ad expenditure}, respectively, using $\delta = 0.4$ and $k = 6$. Market share is expressed as a percentage.
\end{table}%

 Panel B shows that Nongshim products are generally more popular than those of Ottogi, Samyang, and Paldo; Nongshim's average per-product sales are 30\%--50\% higher than those of the other three firms. Ottogi's seven products were sold at the lowest price on average (830 Korean Won), while each was advertised the most. One might argue that Ottogi's favorable consumer reputation and high image score could be partially attributed to its lower price level. Since our focus is on the impact of the corporate image arising exclusively from CSR activities on consumer choice and product sales, we manually reviewed all news articles to ensure that only those highlighting firms’ CSR efforts were included, as detailed in Appendix~A.

\section{Instant noodle demand estimation}\label{section: demand estimation}

 \subsection*{The model}

 In this study, consumer $i$’s indirect utility from product $j$ in market $m$, $u_{ijm}$, is given by:
    \begin{equation}\label{eqn: utility}
    u_{ijm} = \alpha p_{jm} + \beta_1 imgscore_{jt} + \beta_2 cumadv_{jt} + \bold{x}_{jm}\gamma + \xi_{jm} + \nu_{ijm},
    \end{equation}
 where $p_{jm}$ is the product price, and its coefficient $\alpha$ represents the marginal utility of income in absolute terms. Variables $imgscore_{jt}$ and $cumadv_{jt}$ denote the firm-level corporate image score and brand-level cumulative advertising expenditure at time~$t$, respectively, constructed according to equations~\eqref{eqn: image score} and~\eqref{eqn: ad expenditure}. Our main interest lies in the coefficient $\beta_1$; if greater media exposure of a firm's social contributions leads to increased consumer utility from its products, then this coefficient would take a positive value. The vector $\bold{x}_{jm}$ contains a set of dummy variables for product, time, and region, thereby controlling for constant product characteristics, unobserved time-varying demand and supply shocks, and regional differences in consumer tastes. Given the sales seasonality of cold noodle products, the vector also includes interaction terms between the cold noodle indicator and month dummies. $\xi_{jm}$ is the unobserved product and market-specific demand shock.

 We adopt a one-level nested logit specification for instant noodles demand, defining the consumer's idiosyncratic random taste, $\nu_{ijm}$, as
    \begin{equation*}
    \nu_{ijm} = \zeta_{igm} + (1-\sigma)\epsilon_{ijm}, \quad 0 \leq\sigma\leq 1,
    \end{equation*}
 where subscript $g$ represents the product group. As discussed in the previous section, products are partitioned into four groups: red soup, white soup, soupless, and cold noodles ($g=1,\ldots,4$), while the outside good constitutes another group ($g=0$). Assuming that $\epsilon_{ijm}$ is i.i.d. extreme value distributed, $\zeta_{igm}$ has a unique distribution such that $\nu_{ijm}$ is also extreme value distributed \citep{Cardell97}. The nesting parameter $\sigma$ can take any value between 0 and 1; a higher value of $\sigma$ implies stronger substitution between products within the same group. If $\sigma=0$, then there is no additional within-group substitution, and the model collapses to a standard logit.

 Under the one-level nested logit framework, the product's within-group market share, $s_{j|g,m}$, and the market share of the group, $s_{gm}$, are derived as:
    \begin{equation}\label{eqn: market share1}
    s_{j|g,m} = \frac{\exp\Big(\frac{\delta_{jm}}{1-\sigma}\Big)}{\sum_{k\in\mathscr{J}_{gm}}\exp\bigg(\frac{\delta_{km}}{1-\sigma}\bigg)}, \qquad
    s_{gm} = \frac{\left(\sum_{k\in\mathscr{J}_{gm}}\exp\bigg(\frac{\delta_{km}}{1-\sigma}\bigg)\right)^{1-\sigma}}{1+\sum_{g=1}^{4}\left(\sum_{k\in\mathscr{J}_{gm}}\exp\bigg(\frac{\delta_{km}}{1-\sigma}\bigg)\right)^{1-\sigma}},
    \end{equation}
 where $\delta_{jm} = \alpha p_{jm} + \beta_1 imgscore_{jt} + \beta_2 cumadv_{jt} + \bold{x}_{jm}\gamma + \xi_{jm}$ is product $j$'s mean utility in market $m$. The product market share, $s_{jm}$, is the product of the two market shares above:
    \begin{equation}\label{eqn: market share2}
    s_{jm} = s_{j|g,m} \cdot s_{gm}
    \end{equation}

 We then derive a linear demand model by taking the logarithm of both sides of equation \eqref{eqn: market share2} and rearranging the equation:
    \begin{equation}\label{eqn: linear demand equation}
    \begin{aligned}
    \ln\Big(\frac{s_{jm}}{s_{0m}}\Big) & = \delta_{jm} + \sigma\ln s_{j|g,m} \\
    & = \alpha p_{jm} + \beta_1 imgscore_{jt} + \beta_2 cumadv_{jt} + \bold{x}_{jm}\gamma + \sigma \ln s_{j|g,m} + \xi_{jm}
    \end{aligned}
    \end{equation}

 In addition to the standard nested logit model, we also consider the constant expenditures nested logit (CENL) model proposed by \cite{bjornerstedt2016does}. In our baseline utility specification \eqref{eqn: utility}, the price enters linearly, implying that a consumer is supposed to buy up to one unit of his/her most preferred product. Conversely, the price enters logarithmically in the CENL model, assuming that a consumer spends a fixed share of his/her income on the most preferred product. \cite{bjornerstedt2016does} showed that the constant expenditures specification yields price elasticities that do not increase quasi-linearly with prices, a property appealing for analyzing markets with large price variations.

 The demand model under the CENL framework is analogous to our baseline demand specification \eqref{eqn: linear demand equation}:
     \begin{equation}\label{eqn: linear demand equation_CENL}
    \ln\Big(\frac{\tilde{s}_{jm}}{\tilde{s}_{0m}}\Big) = \alpha \ln p_{jm} + \beta_1 imgscore_{jt} + \beta_2 cumadv_{jt} + \bold{x}_{jm}\gamma + \sigma \ln \tilde{s}_{j|g,m} + \xi_{jm},
    \end{equation}
 where $\tilde{s}_{j}$ and $\tilde{s}_{j|g}$ denote the product's revenue share and within-group revenue share, respectively.\footnote{For complete explanations about the constant expenditures utility specification and detailed derivations, please refer to \cite{bjornerstedt2016does}.} Following a similar approach to those in \cite{bjornerstedt2016does} and \cite{kim2022consumers}, we define the size of each market associated with a region as twice the average monthly spending on all instant noodle products in the region during the sample period. We then compute the revenue share of the outside option $\tilde{s}_{0m}$ accordingly. The average market revenue share of product is 1.4\% and that of outside revenue share is 59.8\%.

 \subsection*{Identification}

 The price may be correlated with the unobservable market-specific demand shock. To address this endogeneity concern, we leverage the panel structure of the sample and use the average price of the product in other regions as the instrumental variable for the price \citep{Hausman96, Nevo00, Nevo01}. While prices of a product are correlated across regions, reflecting common cost shocks, the validity of this instrument relies on the assumption that the market-level demand shock, $\xi_{jm}$, is independent across regions. Our utility specification \eqref{eqn: utility} controls for unobservable factors that may affect consumer utility by including product, time, and region fixed effects. Furthermore, the model includes nationwide advertising expenditures for each brand, thereby controlling for another potential source of cross-sectional correlation in $\xi_{jm}$. We believe that after controlling for these unobserved and observed variables, the demand shock $\xi_{jm}$ is likely to be market-specific and thus uncorrelated across regions.

 In addition, we use the number of rival products in the same group as an exogenous instrumental variable \citep{berry95, bresnahan1997market} for the within-group market share. Under the timing assumption that product entry and exit decisions are made before the realization of a market-specific demand shock, the product counts would be exogenous to the shock. Moreover, as more rival products are introduced into the market, the within-group market share would likely decrease, especially in a highly concentrated market such as the Korean instant noodles market.

 In our model, the market-specific error term is unlikely to be correlated with the advertising expenditure, which reflects a firm's marketing activities at the national level. Therefore, we estimate the model by treating $cumadv$ as an exogenous variable. Subsequently, we assess the robustness of our estimation results by allowing a firm's ad spending to be endogenous.

 \subsection*{Results}

 We first estimate a logit model assuming that $\sigma = 0$, and report the estimation results in column (1) of Table \ref{tab: demand estimation}. The results indicate that the price coefficient $\alpha$ has the expected negative sign, while the coefficients on $imgscore$ and $cumadv$, $\beta_1$ and $\beta_2$, have the expected positive signs. Estimates of the nested logit model, reported in column (3) of the table, are consistent with the logit estimates. Notably, the corporate image coefficient $\beta_1$ is statistically significant and positive, suggesting that a rise in consumer favorability toward a firm due to its social contributions would boost demand for the firm's products. We quantify the sales effect of the corporate image improvement in detail in the next section. Additionally, the results show that the estimated nesting parameter $\sigma$ (0.82) is close to one, indicating that consumers view products within the same group as closer substitutes than those in other groups. Column (5) of Table \ref{tab: demand estimation} reports the estimated utility parameters of the CENL model. The results are consistent with those from the baseline nested logit model and show that both media exposure of CSR activities and advertising expenditures positively affect consumer utility. The Sanderson-Windmeijer (SW) first-stage $F$ statistics are large enough to reject that each endogenous regressor is weakly identified. First-stage regression results, presented in the first two columns of Table~\ref{tab:first stage} in Appendix~D, show that the coefficients on the two exogenous instrumental variables are statistically significant and have the expected signs.

\begin{table}[!t]
\small
  \centering
  \caption{Demand estimation results}
    \begin{tabular}{lcccccccc}
    \toprule
          & \multicolumn{2}{c}{Logit} &       & \multicolumn{2}{c}{Nested Logit} &       & \multicolumn{2}{c}{CENL} \\
\cmidrule{2-3}\cmidrule{5-6}\cmidrule{8-9}    Utility parameter & (1)   & (2)   &       & (3)   & (4)   &       & (5)   & (6) \\
    \hline
          &       &       &       &       &       &       &       &  \\
    $\alpha$ (price) & -1.774*** & -1.772*** &       & -0.578*** & -0.578*** &       & -0.362*** & -0.361*** \\
          & (0.061) & (0.084) &       & (0.074) & (0.073) &       & (0.036) & (0.036) \\
    $\sigma$ ($\ln s_{j|g}$) &       &       &       & 0.819*** & 0.819*** &       & 0.703*** & 0.704*** \\
          &       &       &       & (0.051) & (0.050) &       & (0.044) & (0.044) \\
    $\beta_1$ (imgscore) & 1.412*** & 1.412*** &       & 0.177** & 0.177** &       & 0.372*** & 0.371*** \\
          & (0.064) & (0.057) &       & (0.083) & (0.082) &       & (0.070) & (0.070) \\
    $\beta_2$ (cumadv) & 1.212*** & 1.212*** &       & 0.283*** & 0.283*** &       & 0.409*** & 0.408*** \\
          & (0.061) & (0.058) &       & (0.067) & (0.067) &       & (0.061) & (0.061) \\
    \midrule
    Fixed effects &       &       &       &       &       &       &       &  \\
    \hspace{0.1in}Product & Yes   & Yes   &       & Yes   & Yes   &       & Yes   & Yes \\
    \hspace{0.1in}Time & Yes   & -     &       & Yes   & -     &       & Yes   & - \\
    \hspace{0.1in}Region & Yes   & -     &       & Yes   & -     &       & Yes   & - \\
    \hspace{0.1in}Market & -     & Yes   &       & -     & Yes   &       & -     & Yes \\
    \hspace{0.1in}Cold $\times$ Month & Yes   & Yes   &       & Yes   & Yes   &       & Yes   & Yes \\
    \midrule
    Observations & 18,960 & 18,960 &       & 18,960 & 18,960 &       & 18,960 & 18,960 \\
    SW F statistic &       &       &       &       &       &       &       &  \\
    \hspace{0.1in}price & 51941.53 & 76448.36 &       & 426.77 & 430.11 &       & 2090.15 & 2113.38 \\
    \hspace{0.1in}$\ln s_{j|g}$ & -     & -     &       & 350.45 & 352.42 &       & 402.69 & 404.96 \\
    \bottomrule
    \end{tabular}%
  \label{tab: demand estimation}%
\tablenotes
The table reports logit, nested logit, and CENL estimates of the instant noodle demand model. Standard errors, clustered by market, are in parentheses. The estimation results are based on a two-step GMM method. In the regression analysis, the units of the corporate image score ($imgscore$) and cumulative ad spending ($cumadv$) are $\frac{1}{100}$ and 10 billion Korean Won, respectively. This adjustment is made to match the decimal place with other coefficients.
\end{table}%

 Using the estimated nested logit and CENL coefficients, we calculate the implied own- and cross-price elasticities for each observation and average them by product group.\footnote{See Appendix~B for the price elasticities under nested logit and CENL specifications.} The top panel of Table \ref{tab: price elasticities} shows that the average own-price elasticities under the nested logit specification range from -2.75 (Soupless) to -2.11 (Cold noodles). While there is no prior empirical literature on instant noodle demand, our estimate of the product's own-price elasticity (-2.53 overall) is moderate in magnitude compared to the elasticities for other fast-moving consumer goods in South Korea, such as beer (-3.49) and white milk (-3.14), as estimated in \cite{kim2023country} and \cite{kim2022consumers}, respectively. This finding is reasonable given the status of instant noodles as a staple item in the country. In addition, the cross-price elasticities are significantly higher between products within the same group (0.113–1.089) than between products in different groups (0.004–0.006), reflecting the high value of the nesting parameter. According to the bottom panel of the table, the CENL specification yields own-price elasticities that are even smaller than those under the nested logit specification.

\begin{table}[!t]
  \centering
  \caption{Price elasticities}
    \begin{tabular}{lccccc}
    \toprule
    Mean elasticity & Red soup & White soup & Soupless  &  Cold noodles & All \\
    \midrule
          &       &       &       &       &  \\
    \multicolumn{6}{c}{\underline{Nested Logit}}               \\
    Own   & -2.568 & -2.165 & -2.747 & -2.111 & -2.526 \bigstrut[t]\\
    Cross: same & 0.113 & 0.850 & 0.467 & 1.089 & 0.153 \\
    Cross: other & 0.004 & 0.006 & 0.006 & 0.006 & 0.005 \\
          &       &       &       &       &  \\
    \multicolumn{6}{c}{\underline{CENL}}                      \\
    Own   & -2.163 & -1.914 & -2.040 & -1.786 & -2.089 \bigstrut[t]\\
    Cross: same & 0.053 & 0.295 & 0.165 & 0.430 & 0.066 \\
    Cross: other & 0.004 & 0.006 & 0.005 & 0.005 & 0.005 \\
          &       &       &       &       &  \\
    \bottomrule
    \end{tabular}%
  \label{tab: price elasticities}%
\tablenotes Elasticities are calculated using estimated baseline demand coefficients reported in column (3) (nested logit) and column (5) (CENL) of Table \ref{tab: demand estimation}.
\end{table}%

 \subsection*{Robustness}

 To assess the robustness of our demand estimates, we first replace the time and region dummies in our demand specifications with market (region-market) dummies, thereby further controlling for unobserved demand shocks commonly affecting products in the same market. Logit, nested logit, and CENL estimates reported in columns (2), (4), and (6), respectively, of Table \ref{tab: demand estimation} are very close to their counterparts. Second, we construct the proxies for corporate image and ad spending with different values of $\delta$ (the depreciation rate) and $k$ (the maximum time horizon to accumulate the variables), or with a linear decaying pattern as described in Appendix~B, re-estimate the demand models, and report the results in Table \ref{tab: different proxies} in Appendix~D. The effects of media exposure of CSR activities and ad spending on consumer utility remain positive and statistically significant, except in the case where $k = 12$, suggesting that these effects fade after several months.

 We also allow the advertising expenditure to be endogenous and use the number of newly launched products by rival firms as an exogenous instrument for $cumadv$. The validity of this instrument relies on the idea that firms may intensify their marketing activities following the introduction of new products by competitors. Estimation results reported in Table \ref{tab: demand estimation endogenous adv} in Appendix~D are qualitatively similar to our main results in Table \ref{tab: demand estimation}.\footnote{First-stage regression results, presented in the last three columns of Table~\ref{tab:first stage} in Appendix~D, show that the coefficients on the three exogenous instrumental variables are statistically significant and have the expected signs.} While the estimated effects of corporate image and advertising expenditure appear stronger, the Hausman test does not reject the exogeneity of advertising expenditure.

 \section{Corporate image and sales revenues}\label{section: counterfactual}

 In the previous section, we found evidence that improvements in Ottogi's corporate image, resulting from the firm’s diverse CSR activities, have positively affected consumers' utility from its products. Here, we examine the extent to which Ottogi has capitalized on these increasingly favorable perceptions by conducting a series of counterfactual analyses.

 \subsection*{Design of the counterfactual analysis}

 As demonstrated in Section 3, firms in this industry have been subject to \textit{de facto} price regulation, which suggests they have limited ability to optimize product prices. Taking this into account, we assume that prices would have remained unaffected by each firm's relative favorability and treat the observed prices as counterfactual prices. We also derive the counterfactual market shares by simulating prices under the assumption of Bertrand competition, as detailed in Appendix~C.

\begin{figure}[!t]
    \centering
    \caption{Corporate image trends of Ottogi}
    \includegraphics[width=.6\textwidth]{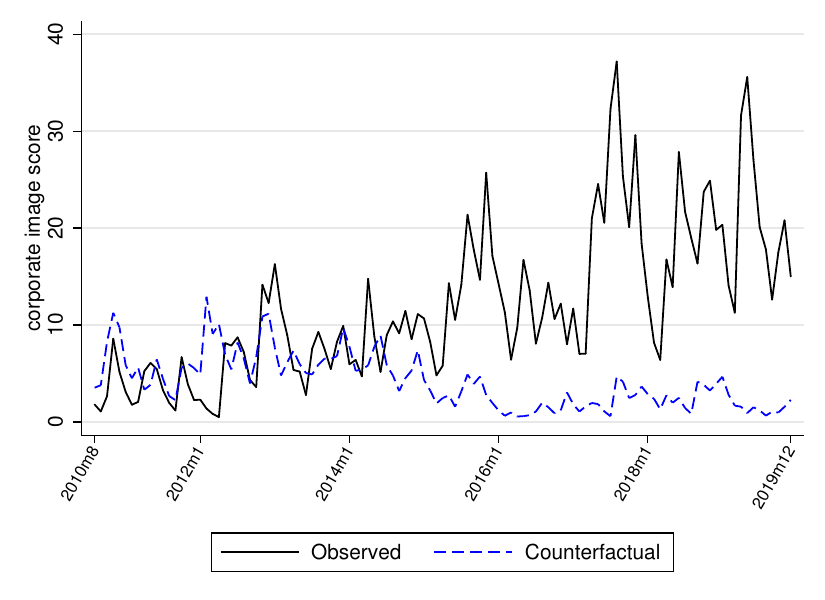}
    \label{fig: simulated image}%
    \tablenotes The black solid line and blue dashed line trace out the observed and counterfactual corporate image scores of Ottogi.
\end{figure}

 We assume that the counterfactual corporate image scores in the absence of Ottogi's image improvement are equal to the observed scores for Ottogi's three competitors. Conversely, we consider the average corporate image score of the three firms at a given time as Ottogi's counterfactual corporate image score for that period. Figure \ref{fig: simulated image} depicts the trends of Ottogi's observed and counterfactual corporate image scores. As demonstrated in Section 3, Ottogi has outperformed its rivals in CSR activities since the mid-2010s. Accordingly, the firm's observed corporate image scores began diverging from the counterfactual score in 2015.

 Using the baseline nested logit estimates presented in column (3) of Table \ref{tab: demand estimation} along with the counterfactual corporate image scores, we simulate sales and revenues of each firm under the scenario in which Ottogi's image evolved similarly to its rivals. Specifically, we calculate the counterfactual mean utility, $\delta^{CF}$:
\begin{equation}\label{eqn: counterfactual mean utility}
\delta_{jm}^{CF} = \hat{\alpha}p_{jm} + \hat{\beta_1}imgscore_{jt}^{CF} + \hat{\beta_2}cumadv_{jt} + \bold{x}_{jm}\hat{\gamma} + \hat{\xi}_{jm},
\end{equation}
where $imgscore_{jt}^{CF}$ is the counterfactual corporate image score constructed above.

 We then insert the simulated mean utility, $\delta_{jm}^{CF}$, into the market share formula given in equation \eqref{eqn: market share1} to obtain the counterfactual within-group market share of the product and the market share of the group, $s_{j|g,m}^{CF}$ and $s_{gm}^{CF}$. Subsequently, the product's counterfactual market share, sales volume, and sales revenue, $s_{jm}^{CF}$, $vol_{jm}^{CF}$, and $rev_{jm}^{CF}$, are calculated as:
\begin{equation}\label{eqn: simulated market share}
	s_{jm}^{CF} = s_{j|g,m}^{CF}s_{gm}^{CF}, ~~ vol_{jm}^{CF} = s_{jm}^{CF}M_{m}, ~~ rev_{jm}^{CF} = vol_{jm}^{CF}p_{jm},
\end{equation}
respectively. By comparing the counterfactual sales and revenues with the observed values, we quantify the sales effects of the corporate image.

 In addition, we estimate how much Ottogi should have increased its advertising spending without the benefit of its corporate image improvement to attain sales revenue equivalent to the actual one. Specifically, we assume a scenario in which the change in Ottogi's image had followed a pattern similar to that of its rivals, but the firm's advertising expenditures had been higher by $\tau$\% than the actual values during the sample period. For different values of $\tau$, we simulate mean utility, $\delta^{CF2}$, using the counterfactual image scores:
 \begin{equation}\label{eqn: counterfactual mean utility2}
\delta_{jm}^{CF2} = \hat{\alpha}p_{jm} + \hat{\beta_1}imgscore_{jt}^{CF} + \hat{\beta_2}(1+\tau)cumadv_{jt} + \bold{x}_{jm}\hat{\gamma} + \hat{\xi}_{jm},
\end{equation}
 which is then used to simulate product sales and revenues. We find the value of $\tau$ that matches the simulated total revenue of Ottogi to its actual total revenue during the sample period.

 \subsection*{Results}

 The top panel of Table~\ref{tab: counterfactual results} presents each firm's actual and simulated sales volumes and revenues during the sample period (August 2010--December 2019), under the assumption of price control by the Korean government. Ottogi's actual sales volume of 3.31 billion packages is approximately 220 million packages (2 million packages per month) higher than the simulated volume of 3.09 billion packages. This finding implies that favorable perceptions of Ottogi may have led to a 6.7\% increase in the firm's sales. Conversely, the sales of other firms were negatively affected by 1.1\%--2.1\% due to Ottogi’s improved image, as consumers substituted their products with Ottogi’s. As a result, Ottogi’s enhanced corporate image led to a 6.5\% increase in its sales revenue and a 1.3\%--2.1\% decrease in revenues for its competitors. Actual industry-wide sales are nearly identical to the simulated values, reflecting strong substitution between products within the same group ($\sigma > 0.8$). According to the results in the bottom panel of the table, Ottogi's social contributions, through its improved corporate image, resulted in 4.1\% and 4.3\% increases in sales and revenue, respectively, had the firms competed \textit{\`a la} Nash-Bertrand. Again, these gains came at the expense of reduced sales for its competitors.

\begin{table}[!t]
  \centering
  \caption{Observed and simulated sales volumes and revenues}
    \begin{tabular}{lrrrrrrr}
    \hline
          & \multicolumn{3}{c}{Sales volumes} &       & \multicolumn{3}{c}{Revenues}  \bigstrut\\
\cline{2-4}\cline{6-8}    Firm  & Observed & Simulated & \% gap &       & Observed & Simulated & \% gap \bigstrut\\
    \hline
          &       &       &       &       &       &       &  \bigstrut[t]\\
    \multicolumn{8}{c}{\underline{Under price regulation}}      \bigstrut[b]  \\
    Ottogi & 3.31  & 3.09  & 6.73  &       & 2.27  & 2.12  & 6.51 \\
    Nongshim & 13.09 & 13.36 & -2.06 &       & 10.40 & 10.62 & -2.08 \\
    Samyang & 2.02  & 2.06  & -1.89 &       & 1.60  & 1.63  & -1.94 \\
    Paldo & 1.07  & 1.08  & -1.13 &       & 0.93  & 0.95  & -1.27 \bigstrut[b]\\
    \hdashline[1pt/1pt]
    Total & 19.50 & 19.59 & 0.00  &       & 15.21 & 15.32 & -0.01 \bigstrut[t]\\
          &       &       &       &       &       &       &  \\
    \multicolumn{8}{c}{\underline{Under Bertrand competition}}  \bigstrut[b]  \\
    Ottogi & 3.31  & 3.18  & 4.07  &       & 2.27  & 2.17  & 4.34 \\
    Nongshim & 13.09 & 13.20 & -0.80 &       & 10.40 & 10.62 & -2.13 \\
    Samyang & 2.02  & 2.07  & -2.51 &       & 1.60  & 1.65  & -3.01 \\
    Paldo & 1.07  & 1.09  & -1.74 &       & 0.93  & 0.95  & -2.09 \bigstrut[b]\\
    \hdashline[1pt/1pt]
    Total & 19.50 & 19.53 & 0.00  &       & 15.21 & 15.40 & -0.01 \bigstrut[t]\\
          &       &       &       &       &       &       &  \bigstrut[b]\\
    \hline
    \end{tabular}%
  \label{tab: counterfactual results}%
\tablenotes The top panel shows the actual and counterfactual total sales volumes and revenues of each firm from August 2010 to December 2019 under the assumption of price regulation, while the bottom panel presents the corresponding figures under the assumption of Bertrand competition. The sales volume is measured in billion packages, and the revenue is in trillion Korean Won.
\end{table}%

\begin{figure}[!t]
\centering
\caption{Revenue trends of Ottogi}
   	\includegraphics[width=.7\textwidth]{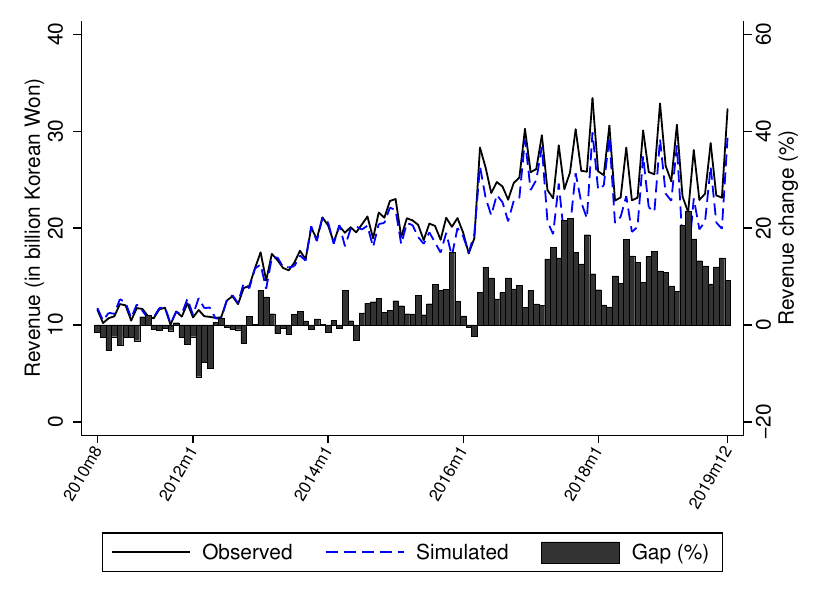}
\label{fig: counterfactual trend}%
\tablenotes The figure depicts trends in Ottogi's actual (black solid line) and simulated (blue dashed line) revenues over the sample period (August 2010 to December 2019). The bar graph illustrates the monthly gap between the actual and simulated revenues.
\end{figure}

 We conduct the following additional analyses under the assumption of price regulation. Figure~\ref{fig: counterfactual trend} illustrates the monthly differences between Ottogi's actual and simulated sales from August 2010 to December 2019, along with their respective trends. The two revenue series largely mirror each other until mid-2014, as Ottogi's corporate image and reputation did not differ significantly from those of its competitors during that period. However, from August 2014 onward, the actual monthly revenue begins to exceed the simulated value by an average of 2.3 billion Korean Won (2.6 million US dollars). This sales uplift coincides with Ottogi's establishment of a distinguished corporate image and reputation among consumers, driven by its diverse CSR initiatives, which began receiving increased recognition in news articles and on social media from the mid-2010s (as shown in Figure~\ref{fig: simulated image}). Figure~\ref{fig: counterfactual other firms} in Appendix~D shows that, in contrast, Ottogi's three competitors lost nearly 3\% of their monthly revenues on average starting in August 2014.

 A comparison of Ottogi's actual and counterfactual sales share trends, plotted in Figure~\ref{fig: counterfactual sales share} in Appendix~D, reveals that Ottogi's sales share in the absence of its distinctive CSR performance (19.2\% on average between August 2014 and December 2019) might have been 2.1 percentage points lower than the actual level (21.3\% on average during the same period). This finding illustrates that the steady increase in Ottogi's market share in the instant noodles market during the 2010s, shown in the top panel of Figure~\ref{fig: share and image}, can be partially attributed to the positive effect of the firm's diverse CSR activities on its corporate image.

\begin{figure}[!t]
\centering
\caption{Simulated revenues of Ottogi under different advertising strategies}
   	\includegraphics[width=.6\textwidth]{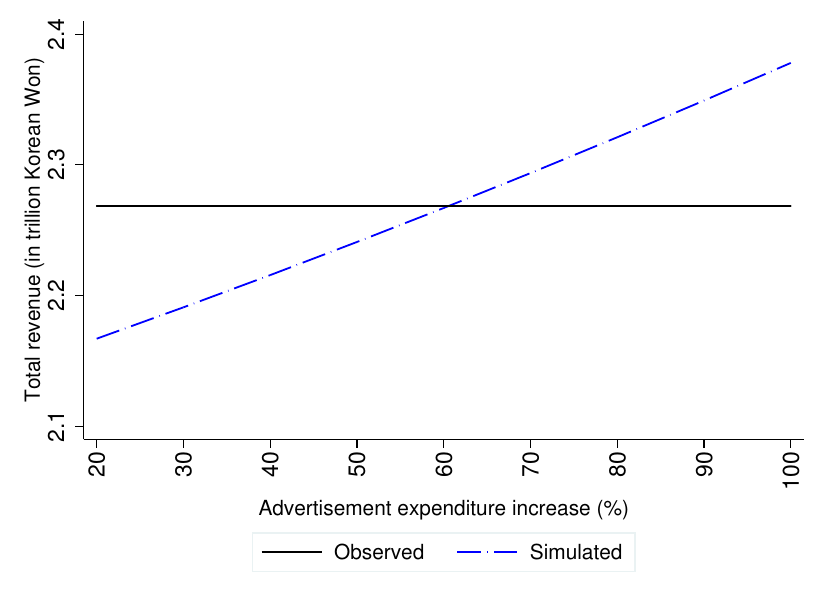}
\label{fig: ad expenditure change}%
\tablenotes In the figure, the black solid line and blue dashed line represent Ottogi's actual and counterfactual revenues, respectively, during the sample period (August 2010 to December 2019). The counterfactual revenues are calculated under various scenarios in which advertising expenditures increase by 20\% to 100\% in the absence of the firm's corporate image improvement.
\end{figure}

 In Figure \ref{fig: ad expenditure change}, we plot Ottogi's predicted sales revenues had the firm spent more on advertising without the benefit of its corporate image improvement (represented by the dotted blue line), alongside its actual revenue during the sample period (shown by the horizontal black line). The figure shows that the revenue effect of corporate image improvement is approximately equivalent in magnitude to the revenue effect of increasing the brand-level advertising expenses by 60\% throughout the sample period. Given that Ottogi spent 78.7 billion Korean Won (70 million US dollars) on advertising during the sample period, this finding implies that the firm should have spent an additional 47.2 billion Korean Won (42 million US dollars) on advertising to achieve the actual revenue of 2.3 trillion Korean Won (2 billion US dollars), without the increasingly positive views of the firm.

 In addition to the baseline counterfactual scenario in which Ottogi's image score follows the average score of its rival firms, we also consider three alternative scenarios where Ottogi’s image score is replaced by that of Nongshim, Samyang, and Paldo, one at a time. Figure~\ref{fig: robust counterfactual} in Appendix~D highlights the rise in Ottogi's image score relative to those of its competitors, which results in a 4.4\%--8.3\% increase in sales volume and a 4.1\%--8.2\% increase in sales revenue during the sample period, as shown in Table~\ref{tab: robust counterfactual}. Moreover, Figure~\ref{fig: ad expenditure change robust} demonstrates that, in the absence of corporate image improvement, Ottogi would have needed to increase advertising expenditures by 38\%--75\% under these scenarios to attain its actual revenue. In summary, we find evidence that engaging in diverse CSR activities and strengthening corporate reputation pays off, and the benefits are substantial.

\section{Conclusion}

 Previous studies have demonstrated that corporate image affects consumers' purchase intentions. Moreover, a deteriorated corporate image reduces consumer utility from the firm's products, leading to a decrease in sales. This negative effect can be long-lasting, as shown by \cite{kim2021corporate}. However, the tangible impact of corporate image improvement remains poorly understood. This issue is particularly important given the increasing number of firms in recent years that have sought to enhance their corporate image through various social contributions. In this study, we empirically examine whether improving corporate image affects consumer choices and firm sales, ands if so, to what extent.

 We focus on the instant noodles industry in South Korea during the 2010s, where Ottogi, one of the four major firms in the industry, established a distinctive and positive corporate image through varied and consistent CSR activities. Utilizing sales and price data from NielsenIQ and a corporate image score constructed from CSR-related news articles, we estimate demand for instant noodles using a nested logit model. Our results suggest robust positive effects of a favorable corporate image on consumers' indirect utility from the firm's products.

 To quantify the sales effects of the firm’s commitment to CSR activities and the resulting improvement in its corporate image, we consider a counterfactual scenario in which the observed change in Ottogi’s image score is replaced by the average of its three competitors. We then predict each product's price, sales volume, and revenue under this scenario and compare these simulated figures to the actual data. Our results suggest a 6.7\% increase in total sales and a 6.5\% increase in total revenues for Ottogi during the sample period (August 2010--December 2019). Additionally, our calculations based on the demand estimation results indicate that a 60\% increase in advertising expenses across all of Ottogi's brands during the sample period would have been necessary to match the actual sales revenue in the absence of the rise in favorability toward the firm.

 In summary, our findings suggest that improvements in corporate image have substantial positive effects on consumer utility and firm sales, providing a practical incentive for firms to actively engage in CSR activities. Such efforts may enhance corporate reputation, which, in turn, can strengthen a firm's competitiveness by fostering greater consumer loyalty to its products.

\section*{Declarations}
\noindent \textbf{Conflicts of interest:} The authors declare that they have no conflict of interest. \\
\noindent \textbf{Data availability statement:} The paper utilizes proprietary data acquired from NielsenIQ and Nielsen Media. Consequently, individuals wishing to replicate the empirical results presented in this paper may need to subscribe to the data services offered by these companies. \\
\noindent \textbf{Code availability:} Matlab and STATA codes are available upon request.

\newpage
\singlespace
\bibliographystyle{econometrica}
\bibliography{reference}

@article{bjornerstedt2016does,
  title={Does merger simulation work? Evidence from the Swedish analgesics market},
  author={Bj{\"o}rnerstedt, Jonas and Verboven, Frank},
  journal={American Economic Journal: Applied Economics},
  volume={8},
  number={3},
  pages={125--164},
  year={2016},
  publisher={American Economic Association 2014 Broadway, Suite 305, Nashville, TN 37203-2425}
}

@article{bresnahan1997market,
  title={Market Segmentation and the Sources of Rents from Innovation: Personal Computers in the Late 1980s},
  author={Bresnahan, Timothy F and Stern, Scott and Trajtenberg, Manuel},
  journal={RAND Journal of Economics},
  volume={28},
  number={0},
  pages={S17--S44},
  year={1997},
}

@article{kim2025collusion,
  title={Collusion suspicion and \textit{de facto} price regulation in the Korean instant noodles market},
  author={Kim, In Kyung and Kim, Kyoo il},
  journal={Available at SSRN 4851827},
  year={2025}
}

@article{kim2022no,
  title={No Beer No Friends: Quantifying the Effect of the Beer Boycott},
  author={Kim, In Kyung and Kim, Kyoo il},
  journal={The Journal of Industrial Economics},
  volume={70},
  number={3},
  pages={711--751},
  year={2022},
}

@article{kim2023country,
  title={Country Image and Consumer Choice: The Case of the Beer Market during the COVID-19 Pandemic},
  author={Kim, In Kyung},
  journal={The Journal of Industrial Economics},
  volume={71},
  number={4},
  pages={1090--1120},
  year={2023},
  publisher={Wiley Online Library}
}

@article{clerides2015national,
  title={National sentiment and consumer choice: The Iraq war and sales of US products in Arab countries},
  author={Clerides, Sofronis and Davis, Peter and Michis, Antonis},
  journal={The Scandinavian Journal of Economics},
  volume={117},
  number={3},
  pages={829--851},
  year={2015},
  publisher={Wiley Online Library}
}

@article{sun2020consumer,
  title={Consumer Boycotts, Country of Origin, and Product Competition: Evidence from China’s Automobile Market},
  author={Sun, Qi and Wu, Fang and Li, Shanjun and Grewal, Rajdeep},
  journal={Management Science},
  year={2020},
  publisher={INFORMS}
}

@article{kim2022consumers,
  title={Consumers' preference for downsizing over package price increases},
  author={Kim, In Kyung},
  journal={Journal of Economics \& Management Strategy},
  volume={33},
  number={1},
  pages={25-52},
  year={2024},
  publisher={Wiley Online Library}
}

@article{auger2003will,
  title={What will consumers pay for social product features?},
  author={Auger, Pat and Burke, Paul and Devinney, Timothy M and Louviere, Jordan J},
  journal={Journal of Business Ethics},
  volume={42},
  number={3},
  pages={281--304},
  year={2003},
  publisher={Springer}
}

@INCOLLECTION{Hausman96,
title = {Valuation of new goods under perfect and imperfect competition},
author = {Hausman, Jerry},
year = {1996},
pages = {207-248},
booktitle = {The Economics of New Goods},
publisher = {National Bureau of Economic Research, Inc},
}

@ARTICLE{Nevo00,
  author = "Nevo, Aviv",
  title = "Mergers with Differentiated Products: The Case of the Ready-to-Eat Cereal Industry",
  journal = "RAND Journal of Economics",
  volume = "31",
  number = "3",
  year = "2000",
  pages = "395-421",
}

@ARTICLE{Nevo01,
  author = "Nevo, Aviv",
  title = "Measuring market power in the ready-to-eat cereal industry",
  journal = "Econometrica",
  volume = "69",
  number = "2",
  year = "2001",
  pages = "307-342",
}

@ARTICLE{Cardell97,
  author = "Cardell, Scott N.",
  title = "Variance components structures for the extreme-value and logistic distributions with application to models of heterogeneity",
  journal = "Econometric Theory",
  volume = "13",
  number = "2",
  year = "1997",
  pages = "185-213",
}

@ARTICLE{Berry95,
  author = "Berry, Steven and Levinsohn, James and Pakes, Ariel",
  title = "Automobile Prices in Market Equilibrium",
  journal = "Econometrica",
  volume = "63",
  number = "4",
  year = "1995",
  pages = "841-890",
}

@article{flammer2015does,
  title={Does corporate social responsibility lead to superior financial performance? A regression discontinuity approach},
  author={Flammer, Caroline},
  journal={Management science},
  volume={61},
  number={11},
  pages={2549--2568},
  year={2015},
  publisher={INFORMS}
}

@article{peters2009some,
  title={Some Evidence of the Cumulative Effects of Corporate Social Responsibility on Financial Performance.},
  author={Peters, Richard and Mullen, Michael R},
  journal={Journal of Global Business Issues},
  volume={3},
  number={1},
  year={2009}
}

@article{jeong2018permanency,
  title={Permanency of CSR activities and firm value},
  author={Jeong, Kwang Hwa and Jeong, Seok Woo and Lee, Woo Jae and Bae, Seong Ho},
  journal={Journal of Business Ethics},
  volume={152},
  pages={207--223},
  year={2018},
  publisher={Springer}
}

@article{list2021corporate,
  title={When corporate social responsibility backfires: Evidence from a natural field experiment},
  author={List, John A and Momeni, Fatemeh},
  journal={Management Science},
  volume={67},
  number={1},
  pages={8--21},
  year={2021},
  publisher={INFORMS}
}

@techreport{hedblom2019toward,
  title={Toward an understanding of corporate social responsibility: Theory and field experimental evidence},
  author={Hedblom, Daniel and Hickman, Brent R and List, John A},
  year={2019},
  institution={National Bureau of Economic Research}
}

@article{mohr2005effects,
  title={The effects of corporate social responsibility and price on consumer responses},
  author={Mohr, Lois A and Webb, Deborah J},
  journal={Journal of consumer affairs},
  volume={39},
  number={1},
  pages={121--147},
  year={2005},
  publisher={Wiley Online Library}
}

@article{martinez2013csr,
  title={CSR and customer loyalty: The roles of trust, customer identification with the company and satisfaction},
  author={Mart{\'\i}nez, Patricia and Del Bosque, Ignacio Rodr{\'\i}guez},
  journal={International journal of hospitality management},
  volume={35},
  pages={89--99},
  year={2013},
  publisher={Elsevier}
}

@article{bachmann2023firms,
  title={Firms and collective reputation: a study of the volkswagen emissions scandal},
  author={Bachmann, R{\"u}diger and Ehrlich, Gabriel and Fan, Ying and Ruzic, Dimitrije and Leard, Benjamin},
  journal={Journal of the European Economic Association},
  volume={21},
  number={2},
  pages={484--525},
  year={2023},
  publisher={Oxford University Press}
}

@article{sharpe2021sales,
  title={Sales response to corporate social irresponsibility and the mitigating role of advertising},
  author={Sharpe, Stacey and Hanson, Nicole},
  journal={Management Decision},
  volume={59},
  number={10},
  pages={2456--2472},
  year={2021},
  publisher={Emerald Publishing Limited}
}

@article{kim2021corporate,
  title={Corporate Social Responsibility and Consumer Choice: Lessons from the Milk Boycott},
  author={Kim, In Kyung and Kim, Kyoo il},
  journal={Management Science},
  note={Forthcoming},
  year={2024}
}

@article{bagwell2007economic,
  title={The economic analysis of advertising},
  author={Bagwell, Kyle},
  journal={Handbook of industrial organization},
  volume={3},
  pages={1701--1844},
  year={2007},
  publisher={Elsevier}
}

@article{clarke1976econometric,
  title={Econometric measurement of the duration of advertising effect on sales},
  author={Clarke, Darral G},
  journal={Journal of Marketing Research},
  volume={13},
  number={4},
  pages={345--357},
  year={1976},
  publisher={SAGE Publications Sage CA: Los Angeles, CA}
}

@article{skarlicki2004third,
  title={Third-party reactions to employee (mis) treatment: A justice perspective},
  author={Skarlicki, Daniel P and Kulik, Carol T},
  journal={Research in organizational behavior},
  volume={26},
  pages={183--229},
  year={2004},
  publisher={Elsevier}
}

@article{hoffmann2018under,
  title={Under which conditions are consumers ready to boycott or buycott? The roles of hedonism and simplicity},
  author={Hoffmann, Stefan and Balderjahn, Ingo and Seegebarth, Barbara and Mai, Robert and Peyer, Mathias},
  journal={Ecological Economics},
  volume={147},
  pages={167--178},
  year={2018},
  publisher={Elsevier}
}

@article{inoue2017predicting,
  title={Predicting behavioral loyalty through corporate social responsibility: The mediating role of involvement and commitment},
  author={Inoue, Yuhei and Funk, Daniel C and McDonald, Heath},
  journal={Journal of business research},
  volume={75},
  pages={46--56},
  year={2017},
  publisher={Elsevier}
}

@article{bagnoli2003selling,
  title={Selling to socially responsible consumers: Competition and the private provision of public goods},
  author={Bagnoli, Mark and Watts, Susan G},
  journal={Journal of Economics \& Management Strategy},
  volume={12},
  number={3},
  pages={419--445},
  year={2003},
  publisher={Wiley Online Library}
}

@article{baron2009positive,
  title={A positive theory of moral management, social pressure, and corporate social performance},
  author={Baron, David P},
  journal={Journal of Economics \& Management Strategy},
  volume={18},
  number={1},
  pages={7--43},
  year={2009},
  publisher={Wiley Online Library}
}

@article{nickerson2022impact,
  title={The impact of corporate social responsibility on brand sales: An accountability perspective},
  author={Nickerson, Dionne and Lowe, Michael and Pattabhiramaiah, Adithya and Sorescu, Alina},
  journal={Journal of Marketing},
  volume={86},
  number={2},
  pages={5--28},
  year={2022},
  publisher={SAGE Publications Sage CA: Los Angeles, CA}
}

@article{van2021does,
  title={Does sustainability sell? The impact of sustainability claims on the success of national brands’ new product introductions},
  author={Van Doorn, Jenny and Risselada, Hans and Verhoef, Peter C},
  journal={Journal of Business Research},
  volume={137},
  pages={182--193},
  year={2021},
  publisher={Elsevier}
}

@article{hao2022corporate,
  title={Corporate social responsibility (CSR) performance and green innovation: Evidence from China},
  author={Hao, Jing and He, Feng},
  journal={Finance Research Letters},
  volume={48},
  pages={102889},
  year={2022},
  publisher={Elsevier}
}

@article{mbanyele2022corporate,
  title={Corporate social responsibility and green innovation: Evidence from mandatory CSR disclosure laws},
  author={Mbanyele, William and Huang, Hongyun and Li, Yafei and Muchenje, Linda T and Wang, Fengrong},
  journal={Economics Letters},
  volume={212},
  pages={110322},
  year={2022},
  publisher={Elsevier}
}

@article{schinkel2024corporate,
  title={Corporate social responsibility by joint agreement},
  author={Schinkel, Maarten Pieter and Treuren, Leonard},
  journal={Journal of Environmental Economics and Management},
  volume={123},
  pages={102897},
  year={2024},
  publisher={Elsevier}
}

@article{calveras2016role,
  title={The role of public information in corporate social responsibility},
  author={Calveras, Aleix and Ganuza, Juan-Jos{\'e}},
  journal={Journal of Economics \& Management Strategy},
  volume={25},
  number={4},
  pages={990--1017},
  year={2016},
  publisher={Wiley Online Library}
}


\newpage
\appendix

\onehalfspacing

\section*{Appendix A: The collection of news articles}

 In this section, we provide detailed explanation about the collection of news articles. We focused on news articles published by major national news media, broadcasting companies, and business and economics newspapers throughout the 2010s, while excluding those from local news media.

 Utilizing the search engine tools provided by Bigkinds, we gathered news articles that included at least one of the CSR keywords outlined in the main text. However, we excluded any articles that contained keywords related to the firm’s market performance, such as `stock,' `sales,' `profit,' `revenue,' and `promotion.' For instance, in the initial collection of news articles without this exclusion procedure, we observed that some articles about Ottogi contained brief descriptions of the firm’s general reputation known for CSR activities, while the main content focused exclusively on its rise in stock value or notable increase in sales or profit. Therefore, this exclusion procedure is crucial to avoid potential reverse causality issues (a rise in news article count due to high sales performance) in the demand estimation and empirical analyses presented in the main text.

 To separate news articles for each of the four firms, we considered only those containing the firm's name and excluded any mentioning its rivals. Our final search formula using keywords for gathering news articles about Ottogi is as follows:
\begin{center}
(At least one CSR keyword) AND (Ottogi) NOT (sales OR stock OR profit OR revenue OR promotion OR names of other three rival firms)
\end{center}
The search formulae for the remaining firms are defined analogously. In total, we initially obtained 5472 news articles: 2518 for Nongshim, 1129 for Ottogi, 1412 for Paldo, and 413 for Samyang.

 We further manually checked all the news articles collected to ensure that only those highlighting the firm’s CSR activities were counted, while excluding any irrelevant articles. First, the names of each of the four firms are often homonyms with different meanings in Korean: `Nongshim' meaning farmer’s spirit, and `Paldo' meaning eight provinces in Korea, usually referring to South Korea in general. After reading the contents of the news articles, we dropped those irrelevant to the firms. Second, we focused on selecting articles where the firm’s CSR activities were the main topic, excluding those where these activities were only briefly mentioned in one or two sentences or merely listed among many other companies. Third, articles related to sponsorships for sports leagues or professional teams were excluded, as they are considered marketing activities, rather than being related to CSR.

 The final news article count is 1154: 350 for Nongshim, 564 for Ottogi, 153 for Paldo, and 87 for Samyang. We believe that our finalized lists of news articles, which highlight the firms' CSR activities contributing to a positive corporate image, serve as appropriate data for constructing the corporate image proxy.


\newpage
\section*{Appendix B: Additional formulae}

 In this section, we present formulae for the alternative construction of the image proxy and cumulative advertising expenditures, as well as the own- and cross-price elasticities under both the Nested Logit and Constant Expenditure Nested Logit models.

 \subsection*{The corporate image proxy}

 In addition to the corporate image score and cumulative advertising expenditures constructed according to equations \eqref{eqn: image score} and \eqref{eqn: ad expenditure} in the main text, we also consider an approach adopting a linear decaying pattern, as follows:
\begin{equation}\label{eqn: linear image score}
	imgscore_{ft} = \sum_{n=0}^{k}(1-n\delta)news_{f,t-n}
\end{equation}
\begin{equation}\label{eqn: linear ad expenditure}
	cumadv_{jt}=\sum_{n=0}^{k}(1-n\delta)adv_{j,t-n},
\end{equation}
 where $news$ is the firm-level monthly CSR-related news article count and $adv$ is the brand-level monthly advertising expenditures. The variable $k$ represents the maximum time horizon over which the variables are accumulated, and $\delta$ is the linear depreciation rate. In our robustness checks, we ensure that $k\delta$ is less than 1 to prevent any positive CSR-related news articles from negatively affecting the corporate image score.

\subsection*{Price elasticities}

 Under the Nested Logit model, the estimated price elasticity between goods $j$ and $k$, denoted as $\hat{\epsilon}_{jk}$, is given by
\begin{equation}\label{eqn: NL elasticity}
	\hat{\epsilon}_{jk} = \hat{\alpha}p_k\frac{s_k}{s_j}\Big(\frac{1}{1-\hat{\sigma}}D_{jk}^1-\frac{\hat{\sigma}}{1-\hat{\sigma}}s_{j|g}D_{jk}^2-s_j\Big),
\end{equation}
where $D_{jk}^1 = 1$ if $k = j$ and $0$ otherwise, and $D_{jk}^2 = 1$ if $j$ and $k$ are in the same group and $0$ otherwise. $\hat{\alpha}$ and $\hat{\sigma}$ are the estimated coefficients from equation \eqref{eqn: linear demand equation}.

Under the Constant Expenditure Nested Logit model, the price elasticity is given by
\begin{equation}\label{eqn: CENL elasticity}
	\hat{\epsilon}_{jk} = \hat{\alpha}\frac{\tilde{s}_k}{\tilde{s}_j}\Big(\frac{1}{1-\hat{\sigma}}D_{jk}^1-\frac{\hat{\sigma}}{1-\hat{\sigma}}\tilde{s}_{j|g}D_{jk}^2 - \tilde{s}_j\Big) - D_{jk}^1,
\end{equation}
where $\tilde{s}_j$ and $\tilde{s}_{j|g}$ represent the product's revenue share and within-group revenue share, respectively. Here, $\hat{\alpha}$ and $\hat{\sigma}$ are the estimated coefficients of equation \eqref{eqn: linear demand equation_CENL}. The Constant Expenditure Nested Logit model and the derivation of price elasticities are delineated in \cite{bjornerstedt2016does}.

\newpage

\renewcommand{\thefigure}{C\arabic{figure}}
\renewcommand{\thetable}{C\arabic{table}}
\setcounter{table}{0}
\setcounter{figure}{0}%

\section*{Appendix C: Counterfactual prices and market shares under Bertrand competition}

 Here, we derive counterfactual prices and market shares under the assumption of Bertrand competition. Firm $f$'s profit in a given market with $J$ products is:
\begin{equation*}
\Pi_{f}(p) = \sum_{j \in \mathcal{J}_{f}} (p_{j} - mc_{j}) q_{j}(p),
\end{equation*}
 where $\mathcal{J}_{f}$ is the set of products owned by the firm, \(mc_{j}\) is product \(j\)'s marginal cost, and $p$ is a $J$ by 1 vector of prices. Utilizing actual prices and market shares, $p^{AT}$ and $s^{AT}$, baseline nested logit estimates presented in column (3) of Table \ref{tab: demand estimation}, and the Nash–Bertrand first-order conditions, we recover the marginal costs in the market as:
\begin{equation*}
\widehat{mc} = p^{AT} + \left(\Omega \circ \nabla_p s \right)^{-1}s^{AT},
\end{equation*}
 where $\widehat{mc}$ is a $J$ by 1 vector of marginal costs, $\Omega$ is the $J$ by $J$ ownership matrix,\footnote{\(\Omega(j,k)\) equals one if products \(j\) and \(k\) belong to the same firm and zero otherwise.} and the operator \(\circ\) denotes element-by-element multiplication. $\nabla_p s$ is the $J$ by $J$ matrix of partial derivatives of market share functions with respect to prices, evaluated using the estimated mean utilities.

 We then obtain counterfactual prices in the absence of Ottogi's image improvement by solving the following $J$ equations simultaneously:
\begin{equation*}
p^{CF} = \widehat{mc} - \left(\Omega \circ \nabla_p s \right)^{-1}s(\delta^{CF}),
\end{equation*}
 where the counterfactual mean utility, used in evaluating the Jacobian matrix and predicting the market shares, is defined as:
\begin{equation*}
\delta_{j}^{CF} = \hat{\alpha}p^{CF}_{j} + \hat{\beta_1}imgscore_{j}^{CF} + \hat{\beta_2}cumadv_{j} + \bold{x}_{j}\hat{\gamma} + \hat{\xi}_{j},
\end{equation*}
 where $imgscore_{j}^{CF}$ is the counterfactual corporate image score, defined in Section 5.

 According to Table \ref{tab: markups}, the predicted price for Ottogi, in the absence of the firm's image improvement (Scenario 1), is on average 0.46\% lower than the actual price. In contrast, for the three other firms, it is 0.12\%--1.20\% higher. This finding suggests that Ottogi leveraged consumers' appreciation of the firm's CSR activities to charge higher prices for its products. Conversely, competitors had to offer lower prices to compensate for consumers' diminished appetite for their products due to their relatively weaker corporate images. The table also shows that the effect of an enhanced corporate image on equilibrium prices is similar under the three alternative counterfactual scenarios in which Ottogi's image score is replaced by that of Nongshim (Scenario 2), Samyang (Scenario 3), and Paldo (Scenario 4), respectively.

\newpage
\begin{table}[htbp]
  \centering
  \caption{Counterfactual prices under Bertrand competition}
    \begin{tabular}{lccccc}
    \toprule
          & Observed & \multicolumn{4}{c}{Counterfactual prices under}  \bigstrut\\
\cline{3-6}    Firm  & prices & Scenario 1 & Scenario 2 & Scenario 3 & Scenario 4 \bigstrut\\
    \hline
          & \multicolumn{1}{c}{} & \multicolumn{1}{c}{} & \multicolumn{1}{c}{} & \multicolumn{1}{c}{} & \multicolumn{1}{c}{} \\
    Ottogi & 685.13 & 681.97 & 682.96 & 681.32 & 681.67 \\
    Nongsim & 794.55 & 804.08 & 801.12 & 806.24 & 804.80 \\
    Paldo & 871.47 & 872.53 & 872.45 & 872.58 & 872.55 \\
    Samyang & 793.19 & 796.46 & 796.12 & 796.71 & 796.56 \\
          &       &       &       &       &  \\
    \bottomrule
    \end{tabular}%
  \label{tab: markups}%
\tablenotes The table reports the predicted prices under four scenarios of Ottogi's image improvement, using the baseline nested logit estimates presented in column (3) of Table~\ref{tab: demand estimation} and the Nash–Bertrand first-order conditions for profit maximization described in Appendix~C.
\end{table}%

\newpage

\renewcommand{\thefigure}{D\arabic{figure}}
\renewcommand{\thetable}{D\arabic{table}}
\setcounter{table}{0}
\setcounter{figure}{0}%

\section*{Appendix D: Additional figures and tables}

\begin{table}[htbp]
  \centering
  \caption{First-stage regression results}
    \begin{tabular}{lcccccc}
    \toprule
          & \multicolumn{2}{c}{Baseline model} &       & \multicolumn{3}{c}{Endogenous ad spending} \\
\cmidrule{2-3}\cmidrule{5-7}    Variable & price & $\ln s_{j|g}$ &       & price & $\ln s_{j|g}$ & cumadv \\
    \midrule
          &       &       &       &       &       &  \\
    price IV & \textbf{0.970***} & -1.245*** & \multicolumn{1}{c}{} & \textbf{0.970***} & -1.282*** & -0.035*** \\
          & \textbf{(0.004)} & (0.060) & \multicolumn{1}{c}{} & \textbf{(0.004)} & (0.062) & (0.011) \\
    share IV & 0.001 & \textbf{-0.140***} & \multicolumn{1}{c}{} & 0.001 & \textbf{-0.148***} & -0.006*** \\
          & (0.000) & \textbf{(0.008)} & \multicolumn{1}{c}{} & (0.000) & \textbf{(0.008)} & (0.001) \\
    adv IV &       &       & \multicolumn{1}{c}{} & -0.000 & 0.024* & \textbf{0.013***} \\
          &       &       & \multicolumn{1}{c}{} & (0.001) & (0.013) & \textbf{(0.003)} \\
    imgscore & -0.004 & 1.255*** & \multicolumn{1}{c}{} & -0.004 & 1.121*** & -0.125*** \\
          & (0.003) & (0.058) & \multicolumn{1}{c}{} & (0.003) & (0.063) & (0.015) \\
    cumadv & -0.001 & 1.102*** & \multicolumn{1}{c}{} &       &       &  \\
          & (0.002) & (0.055) & \multicolumn{1}{c}{} &       &       &  \\
    \midrule
    Fixed effects &       &       &       &       &       &  \\
    \hspace{0.1in}Product & Yes   & Yes   &       & Yes   & Yes   & Yes \\
    \hspace{0.1in}Time & Yes   & Yes   &       & Yes   & Yes   & Yes \\
    \hspace{0.1in}Region & Yes   & Yes   &       & Yes   & Yes   & Yes \\
    \hspace{0.1in}Cold $\times$ Month & Yes   & Yes   &       & Yes   & Yes   & Yes \\
    \midrule
    Observations & 18,960 & 18,960 &       & 18,960 & 18,960 & 18,960 \\
    \bottomrule
    \end{tabular}%
  \label{tab:first stage}%
\tablenotes The first two columns report the first-stage regression results from the baseline nested logit demand model, while the next three columns present the results from the nested logit demand model in which advertising expenditure is treated as endogenous. Price IV, share IV, and adv IV denote the average product price in other regions, the number of rival products in the same group, and the number of newly launched products by rival firms, respectively. Standard errors, clustered by market, are reported in parentheses.
\end{table}

\newpage
\begin{landscape}
\begin{table}[htbp]
  \centering
  \caption{Demand estimation with different corporate image proxy and ad expenditure}
    \begin{tabular}{lccccccccc}
    \hline
          & \multicolumn{5}{c}{Non-linear decaying (geometric sum)} & \textbf{} & \multicolumn{3}{c}{Linear-decaying}  \bigstrut\\
\cline{2-6}\cline{8-10}          & (1)   & (2)   & (3)   & (4)   & (5)   &       & (6)   & (7)   & (8) \bigstrut[t]\\
          &       & $\delta=0.2$ & $\delta=0.6$ & $\delta=0.4$ & $\delta=0.4$ &       & $\delta=0.3$ & $\delta=0.15$ & $\delta=0.08$ \\
    Utility parameter & baseline & $k=6$ & $k=6$ & $k=3$ & $k=12$ &       & $k=3$ & $k=6$ & $k=12$ \bigstrut[b]\\
    \hline
          &       &       &       &       &       &       &       &       &  \bigstrut[t]\\
    $\alpha$ (price) & -0.578*** & -0.602*** & -0.563*** & -0.566*** & -0.399*** &       & -0.568*** & -0.595*** & -0.408*** \\
          & (0.074) & (0.081) & (0.070) & (0.071) & (0.069) &       & (0.071) & (0.079) & (0.081) \\
    $\sigma$ ($\ln s_{j|g}$) & 0.819*** & 0.800*** & 0.832*** & 0.827*** & 0.943*** &       & 0.827*** & 0.805*** & 0.938*** \\
          & (0.051) & (0.057) & (0.046) & (0.048) & (0.039) &       & (0.047) & (0.056) & (0.048) \\
    $\beta_1$ (imgscore) & 0.177** & 0.156** & 0.170** & 0.162** & -0.0230 &       & 0.154** & 0.149** & 0.000 \\
          & (0.083) & (0.069) & (0.085) & (0.079) & (0.061) &       & (0.075) & (0.067) & (0.037) \\
    $\beta_2$ (cumadv) & 0.283*** & 0.213*** & 0.335*** & 0.344*** & 0.0710* &       & 0.260*** & 0.201*** & 0.046 \\
          & (0.067) & (0.055) & (0.073) & (0.077) & (0.039) &       & (0.059) & (0.052) & (0.028) \bigstrut[b]\\
    \hline
    Fixed effects &       &       &       &       &       &       &       &       &  \bigstrut[t]\\
    \hspace{0.1in}Product & Yes   & Yes   & Yes   & Yes   & Yes   &       & Yes   & Yes   & Yes \\
    \hspace{0.1in}Time & Yes   & Yes   & Yes   & Yes   & Yes   &       & Yes   & Yes   & Yes \\
    \hspace{0.1in}Region & Yes   & Yes   & Yes   & Yes   & Yes   &       & Yes   & Yes   & Yes \\
    \hspace{0.1in}Cold $\times$ Month & Yes   & Yes   & Yes   & Yes   & Yes   &       & Yes   & Yes   & Yes \bigstrut[b]\\
    \hline
    Observations & 18,960 & 18,960 & 18,960 & 18,960 & 18,024 &       & 18,960 & 18,960 & 18,024 \bigstrut\\
    \hline
    \end{tabular}%
  \label{tab: different proxies}%
\tablenotes The table reports nested logit demand estimates. The variables $imgscore$ and $cadv$ are constructed using different values of \(\delta\) (the depreciation rate) and \(\kappa\) (the maximum time horizon for accumulating the variables). Column (1) reports the estimation results of the baseline specification with \(\delta=0.4\) and \(\kappa=6\), the same as those reported in column (3) of Table \ref{tab: demand estimation}. The first five columns report estimation results where \(imgscore\) and \(cumadv\) are constructed using equations \eqref{eqn: image score} and \eqref{eqn: ad expenditure}, while the next three columns use equations \eqref{eqn: linear image score} and \eqref{eqn: linear ad expenditure}. Standard errors, clustered by market, are reported in parentheses. The estimation results are based on a two-step GMM method. In the regression analysis, the units of the corporate image score (\(imgscore\)) and cumulative ad spending (\(cumadv\)) are \(\frac{1}{100}\) and 10 billion Korean Won, respectively. This adjustment is made to match the decimal place with other coefficients.
\end{table}%
\end{landscape}

\begin{table}[htbp]
\small
  \centering
  \caption{Demand estimation results under endogenous advertising expenditure}
    \begin{tabular}{lcccccccc}
    \toprule
          & \multicolumn{2}{c}{Logit} &       & \multicolumn{2}{c}{Nested Logit} &       & \multicolumn{2}{c}{CENL} \\
\cmidrule{2-3}\cmidrule{5-6}\cmidrule{8-9}    Utility parameter & (1)   & (2)   &       & (3)   & (4)   &       & (5)   & (6) \\
\hline
          &       &       &       &       &       &       &       &  \\
    $\alpha$ (price) & -1.731*** & -1.728*** &  & -0.590*** & -0.589*** &  & -0.345*** & -0.345*** \\
          & (0.082) & (0.100) &  & (0.073) & (0.073) &  & (0.038) & (0.037) \\
    $\sigma$ ($\ln s_{j|g}$) &       &       &  & 0.789*** & 0.789*** &  & 0.678*** & 0.679*** \\
          &       &       &  & (0.056) & (0.055) &  & (0.050) & (0.050) \\
    $\beta_1$ (imgscore) & 2.122* & 2.122* &  & 0.301*** & 0.301*** &  & 0.479*** & 0.478*** \\
          & (1.195) & (1.200) &  & (0.115) & (0.115) &  & (0.111) & (0.110) \\
    $\beta_2$ (cumadv) & 1.520*** & 1.520*** &  & 0.981** & 0.981** &  & 1.052** & 1.051** \\
          & (0.149) & (0.152) &  & (0.445) & (0.444) &  & (0.518) & (0.516) \\
    \midrule
    Fixed effects &       &       &       &       &       &       &       &  \\
    \hspace{0.1in}Product & Yes   & Yes   &       & Yes   & Yes   &       & Yes   & Yes \\
    \hspace{0.1in}Time & Yes   & -     &       & Yes   & -     &       & Yes   & - \\
    \hspace{0.1in}Region & Yes   & -     &       & Yes   & -     &       & Yes   & - \\
    \hspace{0.1in}Market & -     & Yes   &       & -     & Yes   &       & -     & Yes \\
    \hspace{0.1in}Cold $\times$ Month & Yes   & Yes   &       & Yes   & Yes   &       & Yes   & Yes \\
    \midrule
    Observations & 18,960 & 18,960 &       & 18,960 & 18,960 &       & 18,960 & 18,960 \\
    SW F statistic &       &       &       &       &       &       &       &  \\
    \hspace{0.1in}price & 71.78 & 126.11 &       & 258.69 & 260.90 &       & 276.8 & 277.04 \\
    \hspace{0.1in}$\ln s_{j|g}$ & -     & -     &       & 67.9  & 68.28 &       & 67.64 & 68.03 \\
    \hspace{0.1in}cumadv & 18.5  & 49.18 &  & 16.18 & 16.27 &  & 16.23 & 16.32 \\
    \bottomrule
    \end{tabular}%
  \label{tab: demand estimation endogenous adv}%
\tablenotes The table reports logit, nested logit, and CENL estimates of the instant noodle demand model in which advertising expenditure is treated as endogenous. Standard errors, clustered by market, are in parentheses. The estimation results are based on a two-step GMM method. In the regression analysis, the units of the corporate image score ($imgscore$) and cumulative ad spending ($cumadv$) are $\frac{1}{100}$ and 10 billion Korean Won, respectively. This adjustment is made to match the decimal place with other coefficients.
\end{table}%

\newpage
\begin{table}[htbp]
  \centering
  \caption{Counterfactual sales volumes and revenues under alternative scenarios}
    \begin{tabular}{lrrrrrrr}
    \toprule
          & \multicolumn{3}{c}{Sales volumes} &       & \multicolumn{3}{c}{Revenues} \\
\cmidrule{2-4}\cmidrule{6-8}    Firm  & \multicolumn{1}{l}{Actual} & \multicolumn{1}{l}{Predicted} & \multicolumn{1}{l}{\% gap} &       & \multicolumn{1}{l}{Actual} & \multicolumn{1}{l}{Predicted} & \multicolumn{1}{l}{\% gap} \\
    \midrule
    \\
    \multicolumn{8}{c}{\underline{Scenario 2: Following Nongshim's corporate image trend}}       \bigstrut[b] \\
    Ottogi & 3.31  & 3.17  & 4.40  &       & 2.27  & 2.18  & 4.12 \\
    Nongshim & 13.09 & 13.30 & -1.63 &       & 10.40 & 10.57 & -1.65 \\
    Samyang & 2.02  & 2.05  & -1.52 &       & 1.60  & 1.63  & -1.60 \\
    Paldo & 1.07  & 1.08  & -0.92 &       & 0.93  & 0.94  & -1.02 \bigstrut[b]\\
          &       &       &       &       &       &       &  \\
    \multicolumn{8}{c}{\underline{Scenario 3: Following Samyang's corporate image trend}}      \bigstrut[b] \\
    Ottogi & 3.31  & 3.04  & 8.33  &       & 2.27  & 2.08  & 8.16 \\
    Nongshim & 13.09 & 13.40 & -2.36 &       & 10.40 & 10.65 & -2.38 \\
    Samyang & 2.02  & 2.06  & -2.15 &       & 1.60  & 1.64  & -2.17 \\
    Paldo & 1.07  & 1.09  & -1.26 &       & 0.93  & 0.95  & -1.42 \bigstrut[b]\\
          &       &       &       &       &       &       &  \\
    \multicolumn{8}{c}{\underline{Scenario 4: Following Paldo's corporate image trend}}     \bigstrut[b] \\
    Ottogi & 3.31  & 3.07  & 7.32  &       & 2.27  & 2.11  & 7.12 \\
    Nongshim & 13.09 & 13.38 & -2.17 &       & 10.40 & 10.63 & -2.19 \\
    Samyang & 2.02  & 2.06  & -1.99 &       & 1.60  & 1.64  & -2.03 \\
    Paldo & 1.07  & 1.08  & -1.19 &       & 0.93  & 0.95  & -1.34 \bigstrut[b]\\
    \bottomrule
    \end{tabular}%
  \label{tab: robust counterfactual}%
\tablenotes The table reports the actual and simulated sales volumes and revenues of each firm under alternative scenarios of Ottogi's image improvement. Sales volume is in billion packages, and revenue is in trillion Korean Won.
\end{table}%
\clearpage

\newpage
\begin{figure}[htbp]
	\centering
	\caption{Geographical division of South Korea by NielsenIQ}
	\includegraphics[width=0.9\textwidth]{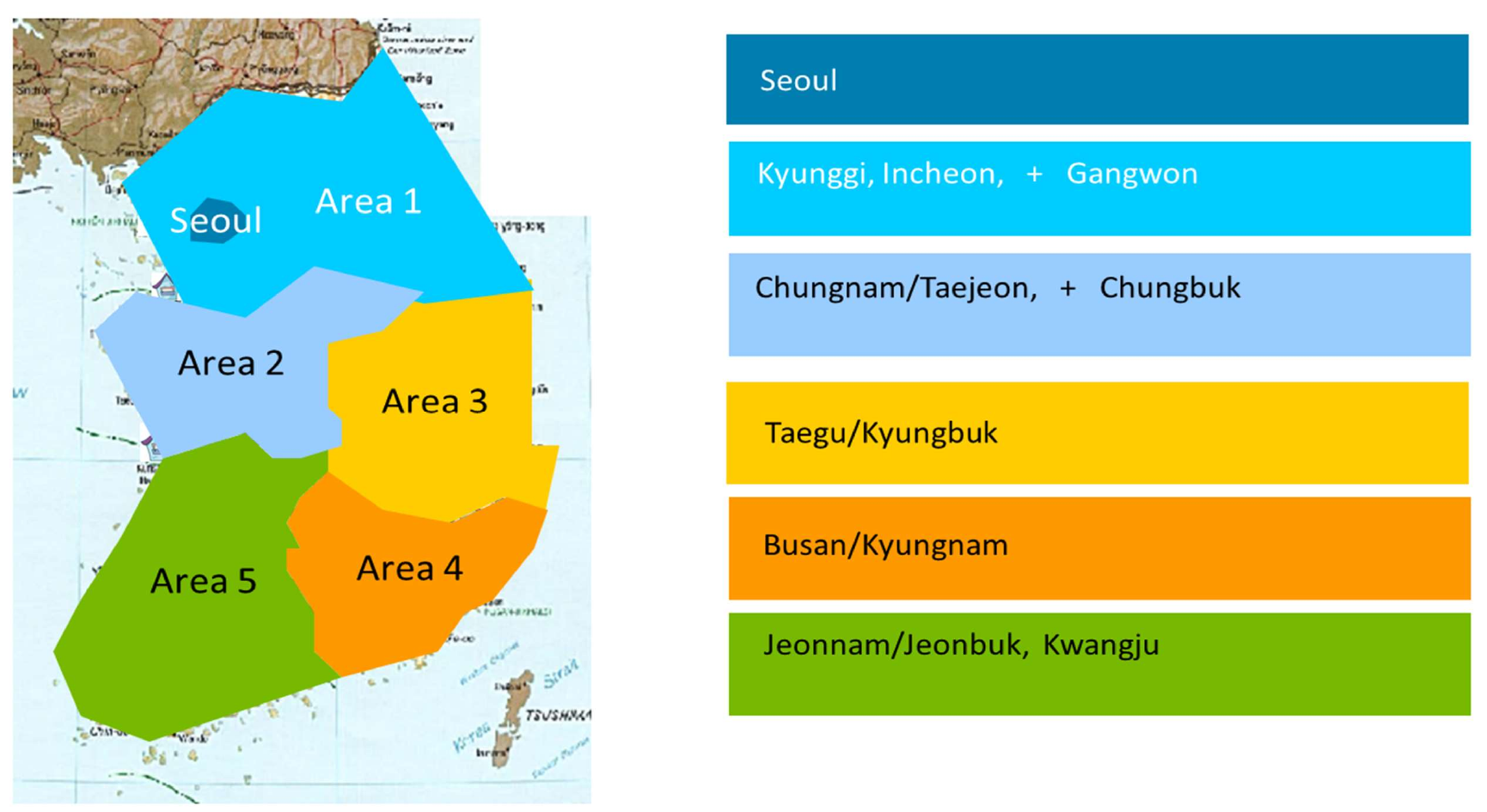}
\label{fig: market nielsen}%
\tablenotes The figure illustrates how NielsenIQ divides South Korea into six geographical markets.
\end{figure}

\newpage
\begin{figure}[htbp]
    \centering
    \caption{Ottogi's corporate image (CI) trends under alternative scenarios}
    \begin{subfigure}[b]{0.5\textwidth}
    \caption{Scenario 2: Following Nongshim's CI trend}
    \includegraphics[width=\textwidth]{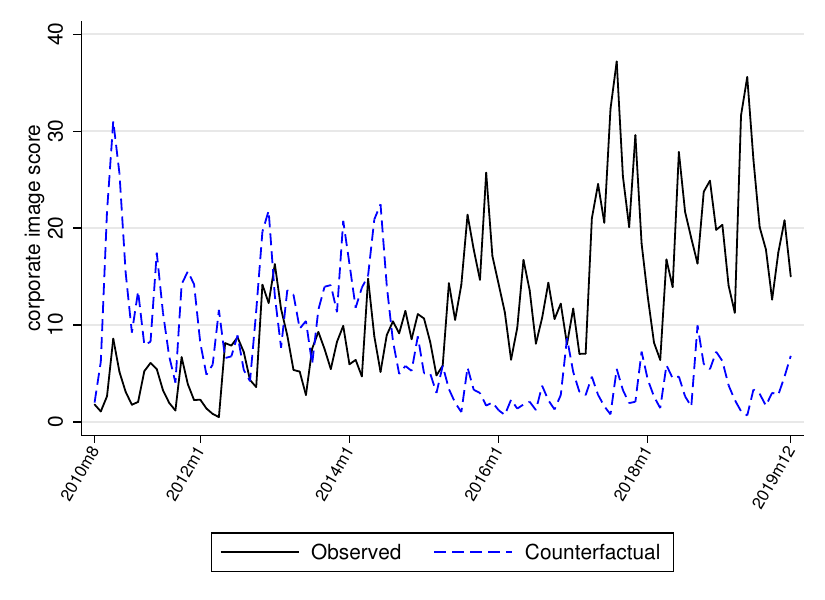}
    \end{subfigure}
    \quad
    \begin{subfigure}[b]{0.5\textwidth}
    \caption{Scenario 3: Following Samyang's CI trend}
    \includegraphics[width=\textwidth]{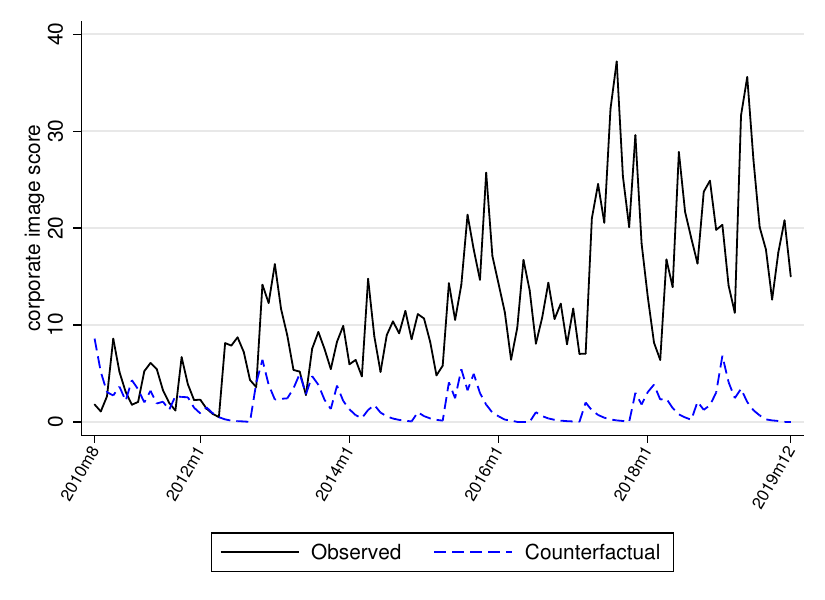}
    \end{subfigure}
    \quad
    \begin{subfigure}[b]{0.5\textwidth}
    \caption{Scenario 4: Following Paldo's CI trend}
    \includegraphics[width=\textwidth]{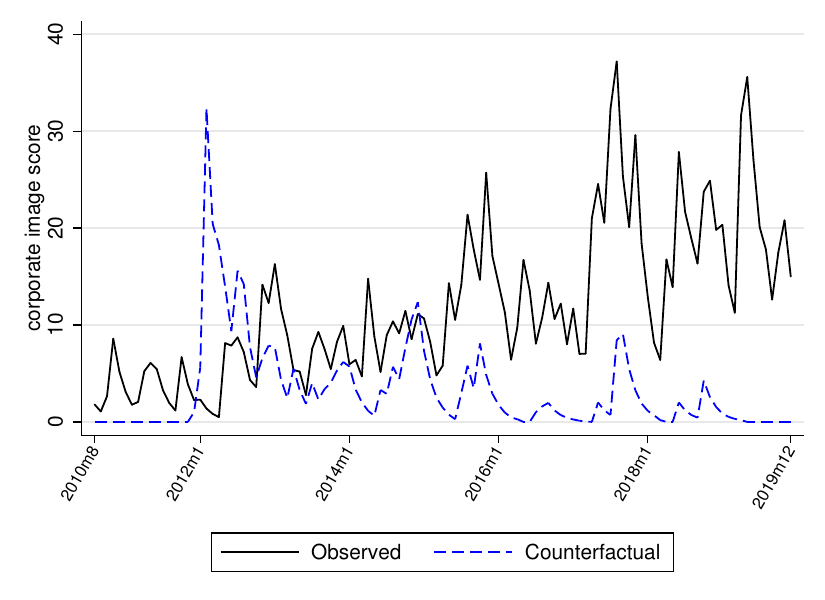}
    \end{subfigure}
\label{fig: robust counterfactual}%
\tablenotes The figure illustrates Ottogi's counterfactual corporate image scores (blue dashed line) under alternative scenarios of image improvement. In Scenarios 2, 3, and 4, Ottogi is assumed to have the same corporate image scores as Nongshim, Samyang, and Paldo, respectively, throughout the sample period.
\end{figure}
\clearpage

\newpage
\begin{figure}[htbp]
    \centering
    \caption{Actual and simulated revenue trends of other firms}
    \begin{subfigure}[b]{0.5\textwidth}
    \caption{Nongshim}
    \includegraphics[width=\textwidth]{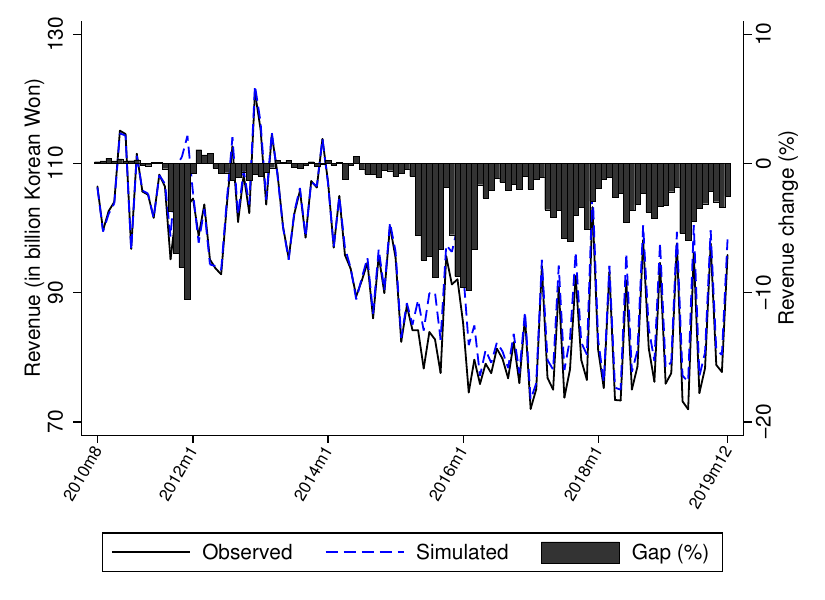}
    \end{subfigure}
    \quad
    \begin{subfigure}[b]{0.5\textwidth}
    \caption{Samyang}
    \includegraphics[width=\textwidth]{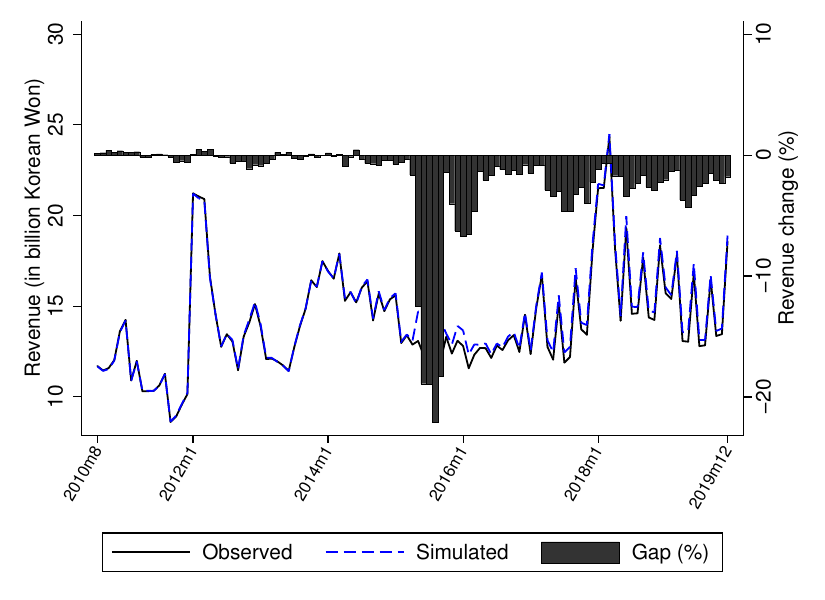}
    \end{subfigure}
    \quad
    \begin{subfigure}[b]{0.5\textwidth}
    \caption{Paldo}
    \includegraphics[width=\textwidth]{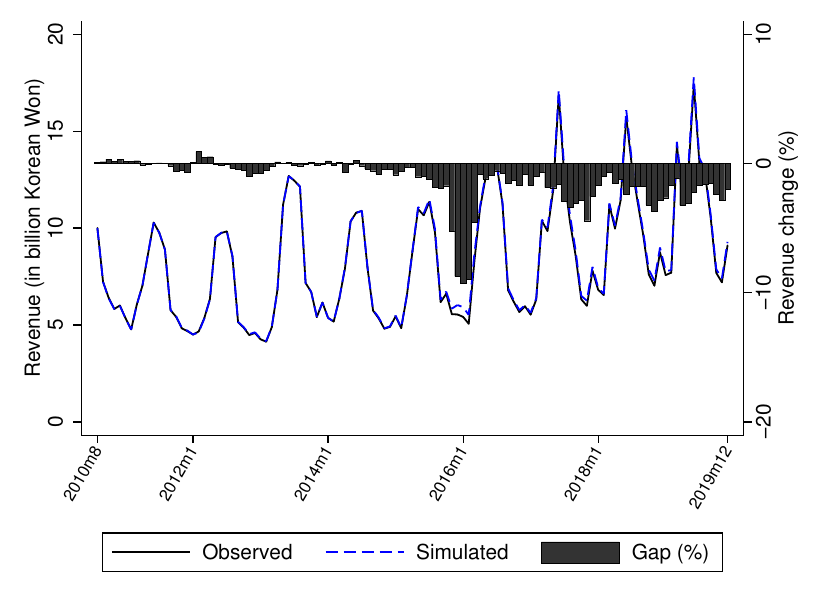}
    \end{subfigure}
\label{fig: counterfactual other firms}%
\tablenotes The figure depicts the trends in actual (black solid line) and simulated (blue dashed line) revenues of Nongshim, Samyang, and Paldo during the sample period (August 2010 to December 2019), under the baseline scenario where Ottogi's corporate image (CI) trend follows the average CI of the three competitors. The bar graph in each panel shows the monthly revenue gap between the actual and simulated revenues.
\end{figure}
\clearpage

\newpage
\begin{figure}[htbp]
	\centering
	\caption{Sales share trends of Ottogi}
	\includegraphics[width=0.65\textwidth]{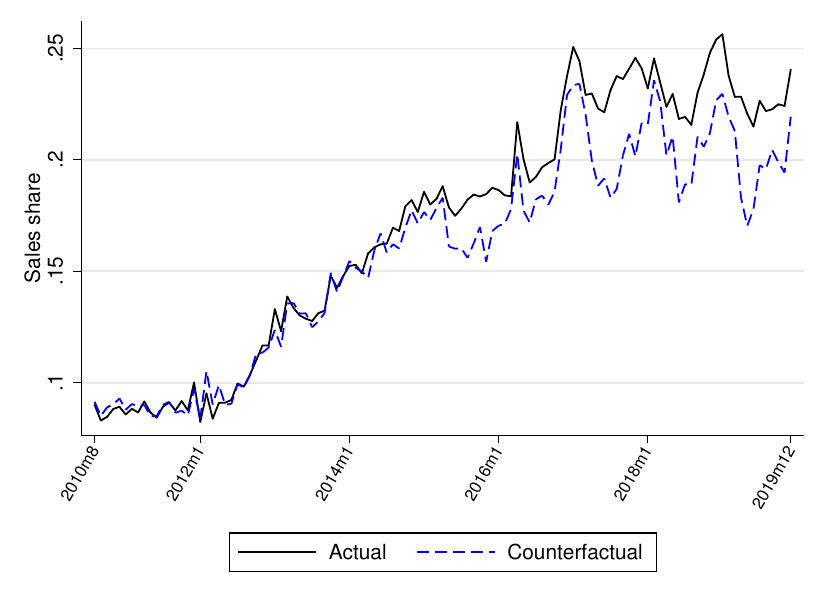}
\label{fig: counterfactual sales share}%
\tablenotes The black solid line and the blue dashed line trace the actual and counterfactual sales shares of Ottogi during the sample period.
\end{figure}


\begin{figure}[!b]
\centering
\caption{Simulated revenues of Ottogi under different advertising and corporate image scenarios}
   	\includegraphics[width=.65\textwidth]{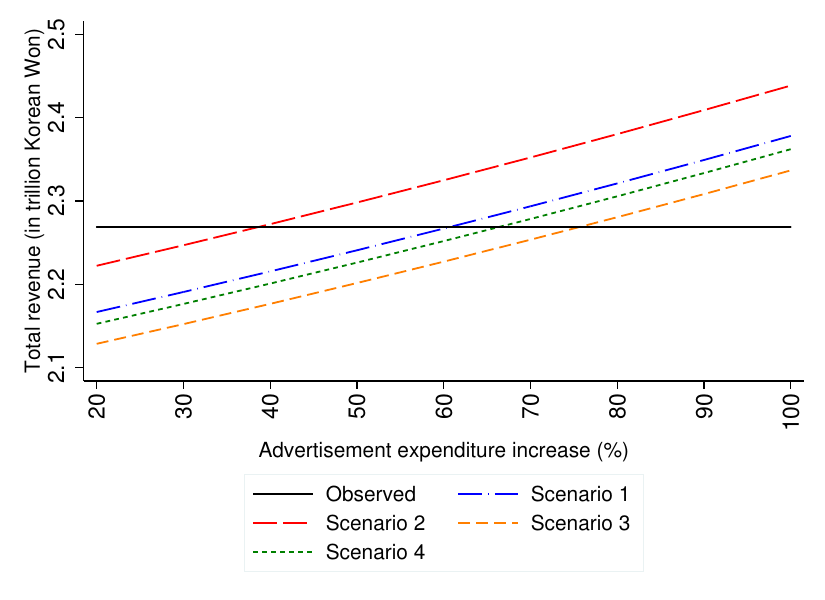}
\label{fig: ad expenditure change robust}%
\tablenotes In the figure, the horizontal black solid line represents Ottogi's observed revenue during the sample period. The four upward-sloping dotted lines depict Ottogi's simulated revenues under four corporate image (CI) trend scenarios: (i) following the average CI trend of its three competitors, (ii) following Nongshim's CI trend, (iii) following Samyang's CI trend, and (iv) following Paldo's CI trend.

\end{figure}

\end{document}